\documentclass[11pt]{article}

\usepackage{graphicx}
\usepackage{amsmath}
\usepackage{amsfonts}
\usepackage{amssymb}
\usepackage{verbatim}

\renewcommand{\arraystretch}{1.3}

\catcode`\@=11
\def\marginnote#1{}

\newcount\hour
\newcount\minute
\newtoks\amorpm
\hour=\time\divide\hour by60
\minute=\time{\multiply\hour by60 \global\advance\minute by-\hour}
\edef\standardtime{{\ifnum\hour<12 \global\amorpm={am}%
        \else\global\amorpm={pm}\advance\hour by-12 \fi
        \ifnum\hour=0 \hour=12 \fi
        \number\hour:\ifnum\minute<10 0\fi\number\minute\the\amorpm}}
\edef\militarytime{\number\hour:\ifnum\minute<10 0\fi\number\minute}

\def\draftlabel#1{{\@bsphack\if@filesw {\let\thepage\relax
      \xdef\@gtempa{\write\@auxout{\string
          \newlabel{#1}{{\@currentlabel}{\thepage}}}}}\@gtempa \if@nobreak
    \ifvmode\nobreak\fi\fi\fi\@esphack} \gdef\@eqnlabel{#1}}
    \def\@eqnlabel{}
\def\@vacuum{}
\def\draftmarginnote#1{\marginpar{\raggedright\scriptsize\tt#1}}

\def\draft{
  \oddsidemargin -.5truein
  \def\@oddfoot{\footnotesize \sl preliminary draft \hfil
    \rm\thepage\hfil\sl\today\quad\militarytime}
  \let\@evenfoot\@oddfoot \overfullrule 3pt
    \let\label=\draftlabel
    \let\marginnote=\draftmarginnote
  \def\@eqnnum{(\theequation)\rlap{\kern\marginparsep\tt\@eqnlabel}%
    \global\let\@eqnlabel\@vacuum}

  }

\makeatletter
\newdimen\normalarrayskip              
\newdimen\minarrayskip                 
\normalarrayskip\baselineskip
\minarrayskip\jot
\newif\ifold             \oldtrue            \def\new{\oldfalse}
\def\arraymode{\ifold\relax\else\displaystyle\fi} 
\def\eqnumphantom{\phantom{(\theequation)}}     
\def\@arrayskip{\ifold\baselineskip\z@\lineskip\z@
     \else
     \baselineskip\minarrayskip\lineskip2\minarrayskip\fi}
\def\@arrayclassz{\ifcase \@lastchclass \@acolampacol \or
\@ampacol \or \or \or \@addamp \or
   \@acolampacol \or \@firstampfalse \@acol \fi
\edef\@preamble{\@preamble
  \ifcase \@chnum
     \hfil$\relax\arraymode\@sharp$\hfil
     \or $\relax\arraymode\@sharp$\hfil
     \or \hfil$\relax\arraymode\@sharp$\fi}}
\def\@array[#1]#2{\setbox\@arstrutbox=\hbox{\vrule
     height\arraystretch \ht\strutbox
     depth\arraystretch \dp\strutbox
     width\z@}\@mkpream{#2}\edef\@preamble{\halign
\noexpand\@halignto
\bgroup \tabskip\z@ \@arstrut \@preamble \tabskip\z@ \cr}%
\let\@startpbox\@@startpbox \let\@endpbox\@@endpbox
  \if #1t\vtop \else \if#1b\vbox \else \vcenter \fi\fi
  \bgroup \let\par\relax
  \let\@sharp##\let\protect\relax
  \@arrayskip\@preamble}
\def\eqnarray{\stepcounter{equation}%
              \let\@currentlabel=\theequation
              \global\@eqnswtrue
              \global\@eqcnt\z@
              \tabskip\@centering
              \let\\=\@eqncr

 \halign to \displaywidth\bgroup
    \eqnumphantom\@eqnsel\hskip\@centering
    $\displaystyle \tabskip\z@ {##}$%
    \global\@eqcnt\@ne \hskip 2\arraycolsep
         $\displaystyle\arraymode{##}$\hfil
    \global\@eqcnt\tw@ \hskip 2\arraycolsep
         $\displaystyle\tabskip\z@{##}$\hfil
         \tabskip\@centering
    &{##}\tabskip\z@\cr}
\newfont{\hr}{msbm10}
\newfont{\ams}{msam10}

\voffset= - 1.2in
\hoffset= - 1.0in         

\def\beq{\begin{equation}}
\def\eeq{\end{equation}}
\def\ba{\beq\new\begin{array}{c}}
\def\ea{\end{array}\eeq}
\def\be{\ba}
\def\ee{\ea}

\def\res{{\rm res}}

\def\F{{\cal F}}

\def\d{\partial}
\def\N2{${\cal N}=2$}

\def\1N{${\cal N}=1$}
\def\4N{${\cal N}=4$}
\def\nn{\nonumber}

\def\p{\partial}

\def\const{\hbox{const}}
\def\res{\hbox{res}}

\renewcommand{\d}{\partial}
\newcommand{\D}{\nabla}

\newcommand{\from}{\geqslant}
\newcommand{\dd}{\tilde{\partial}}
\renewcommand{\ni}{\noindent}
\def\F{\mathcal{F}}
\newcommand{\Rc}{\check{R}}
\newcommand{\aW}{W}
\newcommand{\wt}[1][F]{{\widetilde{#1}}}

\DeclareMathOperator{\rT}{T}
\newcommand{\pq}[2]{\hbox{\tiny{(#1,#2)}}}
\renewcommand{\tt}[1][mer]{\hbox{\tiny{#1}}}
\newcommand{\proof}{\noindent\textbf{Proof\,:\;}}

\newcommand{\cE}{\mathcal{E}}
\newcommand{\cEt}{\cE^\circlearrowleft}
\newcommand{\cG}{\mathcal{G}}
\newcommand{\eqdef}{\stackrel{def}{=}}

\newtheorem{prop}{Proposition}[section]
\newtheorem{lemma}[prop]{Lemma}

\newcommand{\Tr}{\mathop{\rm Tr}\nolimits}

\def\theequation{\arabic{section}.\arabic{equation}}

\title{{\bf
Partition Functions of Matrix Models
\\
as the First Special Functions of String Theory
\\
II. Kontsevich Model} \vspace{.5cm}}
\author{{\bf A. Alexandrov}\thanks{E-mail: \ al@itep.ru}
\date{ } \\ {\small
{\it IHES, 91440 Bures-sur-Yvette, France and ITEP, Moscow, Russia}}\\ \\
{\bf A. Mironov}\thanks{E-mail:
\ mironov@itep.ru; mironov@lpi.ac.ru}
\date{ } \\
{\small {\it Theory Department, Lebedev Physics Institute}
and {\it ITEP, Moscow, Russia}}\\ \\
{\bf A.Morozov}\thanks{E-mail: \ morozov@itep.ru}
\date{ } \\ {\small
{\it ITEP, Moscow, Russia}}\\ \\
{\bf P.Putrov}\thanks{E-mail: \ putrov@gmail.com}
\date{ } \\ {\small
{\it S.-Petersburg State University, S.-Petersburg, Russia}}}

\begin{document}

\setcounter{footnote}{3}

\setcounter{tocdepth}{3}

\maketitle

\vspace{-13.5cm}

\begin{center}
\hfill FIAN/TD-22/08\\
\hfill ITEP/TH-49/08\\
\hfill IHES/P/08/52\\
\end{center}

\vspace{11.5cm}

\begin{abstract}
In arXiv:hep-th/0310113 we started a program of creating a
reference-book on matrix-model $\tau$-functions, the new generation
of special functions, which are going to play an important role in
string theory calculations. The main focus of that paper was on the
one-matrix Hermitian model $\tau$-functions. The present paper is
devoted to a direct counterpart for the Kontsevich and Generalized
Kontsevich Model (GKM) $\tau$-functions. We mostly focus on
calculating resolvents (=loop operator averages) in the Kontsevich
model, with a special emphasis on its simplest (Gaussian) phase,
where exists a surprising integral formula, and the expressions for
the resolvents in the genus zero and one are especially simple (in
particular, we generalize the known genus zero result to genus one).
We also discuss various features of generic phases of the Kontsevich
model, in particular, a counterpart of the unambiguous Gaussian
solution in the generic case, the solution called Dijkgraaf-Vafa
(DV) solution. Further, we  extend the results to the GKM and, in
particular, discuss the $p$-$q$ duality in terms of resolvents and
corresponding Riemann surfaces in the example of dualities between
(2,3) and (3,2) models.
\end{abstract}
\newpage
\tableofcontents

\newpage

\section{Introduction}

\subsection{Aim of the paper}

In \cite{amm1} a program was started to create a reference-book on
matrix-model $\tau$-functions -- the new generation of special
functions, which are going to play an  important role in string
theory calculations. The goal is to extract and considerably extend
spectacular results obtained during the golden era of matrix model
studies in late 80's and during sporadic moments of emerging new
interest afterwards (see, e.g., reviews \cite{80,mamo_r} and
references therein). In \cite{amm2}-\cite{AMM.IM} a number of steps
was made towards realization of this program for the {\it most
fundamental} partition function of Hermitian matrix model.
Integrable aspects of that theory were earlier considered in
\cite{intmamo,mamo_r}. Additional progress is made in the QFT-like
approach, which is being developed in a complementary series of
papers \cite{Eyn}. While we are still far from having a concise and
exhaustive presentation in case of the Hermitian matrix model, time
is also coming to extend analysis to other matrix-model
$\tau$-functions. The present paper being a sequel of \cite{amm1} is
a direct counterpart of \cite{amm1} for the Kontsevich \cite{Ko} and
Generalized Kontsevich Model (GKM) \cite{GKM,versus}
$\tau$-functions. For parallel consideration see \cite{EyKo}. Again,
integrable and QFT aspects of the theory are mainly not included: we
concentrate mostly on Virasoro-like constraints, perturbative
resolvents (multi-densities) and duality properties, which are still
insufficiently represented in the literature. In fact, the situation
with the Kontsevich $\tau$-function {\it per se} is a little better
than with the Hermitian model, because its direct relation to the
topological field theory \cite{top} stimulated a relatively
systematic consideration in the past, see \cite{Itzykson}-\cite{ok}
and references therein. Still, the most interesting part of the
story -- that of $(p,q)$-$(q,p)$ duality -- remained almost
untouched after preliminary papers \cite{FKN,quas,KM}.

As soon as the topic can be hardly exhausted within one paper, here
we concentrate only on a few basic examples leaving further
developments for future publications. In particular, we mostly focus
on calculating resolvents (=loop operator averages) in the
Kontsevich model, with a special emphasis on its simplest (Gaussian)
phase, where exists a surprising integral formula \cite{ok}, and the
expressions for the resolvents in genera zero and one are especially
simple (in particular, we generalize the known genus zero result
\cite{AJKM} to genus one, see also \cite{AK}). Thus, the program of
\cite{amm1} is realized for the Kontsevich model with several
important simplifications: in variance with \cite{amm1}, there are
very simple formulas for the $n$-point resolvent for lower genera;
there is an integral formula for the Laplace transform of the
$n$-point resolvent (summed over all genera!); and the (integral)
equation for the generating function of all resolvents looks
especially simple. In the paper, we give a review of these
simplifications in sect.3, while before that, in sect.2, we discuss
various features of generic phases of the Kontsevich model. Note
that in a generic phase the main method to calculate resolvents is
to solve the loop equations \cite{loopeq,FKN1,DVV}, which generally
have many solutions, the Gaussian Kontsevich model being the only
one that has the {\it unique} solution. However, there is a
counterpart of this {\it unambiguous} Gaussian solution in the
generic case, the solution called Dijkgraaf-Vafa (DV) solution
\cite{DV}-\cite{DVmore}. Among other features, this solution has
specific integrable properties. We discuss the DV solution in
sect.4. The generalization of results to the Generalized Kontsevich
model is contained in sect.5-6, where we also discuss the $p$-$q$
duality in terms of resolvents and corresponding Riemann surfaces in
the example of dualities between (2,3) and (3,2) models.

\subsection{Correlators in matrix models}

By essence, the main problem of matrix models one may address to is
constructing a quantum field theory (QFT) presentation of matrix
models. Solving this problem would allow one to effectively deal
with all other problems.

The main purpose of QFT study of any model is to evaluate arbitrary
correlation functions in an arbitrary phase and, after that, to
study possible relations (``dualities'') between these correlators
in different phases. In the context of matrix models certain subsets
of correlation functions are naturally collected into generating
functions which will be called resolvents (or multi-densities). They
possess, at least, three different representations.

{\it La raison d'etre} for (multi-)resolvents is a transparent
group-theoretical structure of the Schwinger-Dyson equations (which
is obscure in a generic QFT but is immediately obvious in the simple
matrix models): these are $W$- (Virasoro in the simplest cases)
constraints with a loop-algebra structure. Accordingly, correlation
functions satisfy the loop-equations, and the loop parameter becomes
a natural expansion parameter of the generating functions.

It is still difficult to solve the genuine loop equation and obtain
the {\it full} generating function, but an additional ``genus
expansion'' converts the loop equation into a chain of simpler loop
equations for {\it partial} generating functions, multi-resolvents
which can be evaluated straightforwardly one after another.
Ambiguities arising in this recursive process lead to different sets
of multi-resolvents and are interpreted as associated with different
phases of the theory. This will be our first approach to
multi-resolvents.

Multi-resolvents emerge as non-trivial functions of the loop
parameter $z$ with singularities of various types, both poles and
branchings. The second approach deals with them as
poly-differentials on an auxiliary Riemann surface $\Sigma_0$
(``spectral'' complex curve), and different phases correspond to
different choices of the spectral surface and, in addition, to
different conditions on the periods of multi-resolvent
poly-differentials (e.g., if all the periods but the periods of the
first multi-resolvent are vanishing, one gets to the so called
Dijkgraaf-Vafa phase). As usual, poly-differentials on Riemann
surfaces are most immediately represented as correlators of free
fields, hence, this approach is often called conformal field theory
(CFT) representation. An adequate reformulation of the loop
equations suitable for the CFT representation is partly worked out
in \cite{Eyn,AMM.IM}.

The third approach represents correlation functions via (functional
or matrix) integrals. The problem, however, is that the
multi-resolvents are {\it generating functions} of matrix model
correlators, i.e. derivatives of the matrix model partition
function, and, therefore, their representation by integral formulas
is not {\it a priori} obvious. In the Hermitian model this
representation is rather straightforward: the spectral (or loop)
parameter $z$ is introduced through the average of the loop operator
\be \Tr \frac{dz}{z-\phi}=
\sum_{k=0}^\infty\frac{dz}{z^{k+1}}\Tr\phi^k\longrightarrow
\sum_{k=0}^\infty\frac{dz}{z^{k+1}}\frac{\d}{\d t_k}\ee where $\phi$
is the Hermitian matrix that is integrated over. However, a
counterpart representation for the Kontsevich model remains unclear
(see \cite{quas} for a very tedious approach to evaluating a few
first $\frac{\d}{\d t_k}$ for the generalized Kontsevich integral).
Worse than that, even integral formulas for the partition function
are also unknown for most of non-trivial phases of the Kontsevich
model. Still, spectacular results for correlators of the Gaussian
Kontsevich model, due to \cite{ok} (discussed in sect.3), imply that
the third approach should also be fruitful. Somewhat surprisingly
the integral formulas in the Gaussian case are most simple not for
the multi-resolvents themselves (i.e. not for the quantities
subjected to the loop equations), but for their Laplace transforms.

\subsection{IZK integral}

Now we specify our general discussion to the case of the Kontsevich
model, the main object of the present paper.

The story about the Kontsevich model begins from the
Itzykson-Zuber-Kontsevich (IZK) integral over $n\times n$ Hermitian
matrices $X$ \be I(\Lambda|V) = \frac{1}{{\cal N}(\Lambda)}\int dX
\exp \Big({\rm tr}\ \Lambda X\ - \ {\rm tr}\ V(X)\Big), \label{IZ}
\ee depending on the choice of the potential $V(x)$, \be V(x) =
\sum_{k=0}^\infty s_k x^k \ee and on the background matrix-valued
field $\Lambda$. ${\cal N}(\Lambda)$ here is a normalization factor.
This matrix model is actually of {\it the eigenvalue type}
\cite{intmamo,mamo_r}: as was first demonstrated by Itzykson and
Zuber \cite{IZI}, the integration over angular variables $U$ in $X =
U^+X_{diag}U$ can be done explicitly, leaving the $n$-fold integral
over eigenvalues $\chi_i$ of $X$ in $X_{diag} = {\rm
diag}(\chi_1,\ldots, \chi_n)$, \be I(\Lambda|V) \sim \prod_{i=1}^n
\int d\chi_i e^{-V(\chi_i)}\frac{\Delta(\chi)}{\Delta(\lambda)}
\det_{i,j=1,\ldots,n} e^{\chi_i \lambda_j} \label{IZev} \ee where
$\lambda_j$ are eigenvalues of $\Lambda$ and $\Delta$ denotes the
Van-der-Monde determinant,
$$\Delta(\chi)  \equiv
\det_{i,j=1,\ldots,n} \chi_i^{j-1}= \prod_{i<j}^n (\chi_i-\chi_j)$$

Transition from (\ref{IZ}) to (\ref{IZev}) is typical for the
Harish-Chandra-style character calculus in group theory
\cite{HCh,GKM2}. A particular expansion of the particular IZK
integral with pure cubic  potential $V(x) = x^3$ was related by
M.Kontsevich \cite{Ko} to cohomologies of the moduli space of
Riemann surfaces and, finally \cite{MMMtgr}, to partition function
of topological gravity \cite{top}. Moreover, this particular cubic
potential case turns out to be related to more general Hodge
integrals over the moduli space that include $\lambda$-classes
\cite{HodgeK}, these latter being related to the Hurwitz numbers
\cite{ELSV}. Many properties of the integral are, however,
independent on particular choice of $V(x)$ and can be addressed in
the theory of {\it Generalized Kontsevich Model} (GKM)
\cite{GKM,GKMU}.

$\bullet$ The first split between different directions of study of
the GKM concerns the type of $\Lambda$-dependence in (\ref{IZ}). One
option is to consider ${\rm tr}\ \Lambda X$ as a perturbation and
represent $I(\Lambda)$ as a series in powers of ${\rm tr}\
\Lambda^k$ with $k> 0$ -- this is the {\it character phase} of the
model \cite{GKM2,GKMU}. Instead one can expand around a classical
solution $X=L$ to the equation of motion $V'(L) = \Lambda$ of the
full action, then the expansion will be in powers of $t_k =
\frac{1}{k}{\rm tr} L^{-k}$ with $k>0$ provided the normalization
factor ${\cal N}$ in (\ref{IZ}) is chosen equal to the
quasiclassical value of the integral -- this is the {\it Kontsevich
phase} of the same model. In the GKM with {\it monomial}
potential\footnote{Note that $p$ here can be negative equally well,
the anti-polynomial Kontsevich case, \cite{GKMU}.}
$$V_p(x) = \frac{x^{p+1}}{p+1}$$
$L$ is just one of the $p$-th roots of $\Lambda$: $L^p = \Lambda$,
moreover, in this case the integral does not depend on $T_k$ with
$k$ divisible by $p$ \cite{GKM}. For non-monomial potentials $V(x)$
there are essentially different choices of $L$ and essentially
different Kontsevich phases (in their simplest phase, non-monomial
potentials are reduced to the monomial ones, see \cite{quas,Kr}).

\bigskip

$\bullet$ As usual for matrix models, the original integral
(\ref{IZ}) is not an adequate definition of the partition function:
as it is, it describes reasonably only {\it some} of the phases. At
the next step, it should be substituted with a set of differential
equations w.r.t. the time variables $t_k$ and $s_k$ so that the
partition function is defined to be a generic multi-branch solution
to this system, with (\ref{IZ}) providing integral representations
for some of the branches. These equations have a simple form of {\it
continuous Virasoro constraints} for the simplest case of
$V(x)=\frac{1}{3}x^3$ \cite{FKN1,DVV,Wit,MMMtgr}, i.e. for the
original {\it Kontsevich model} \cite{Ko}, become more sophisticated
$W^{(p+1)}$-constraints for the GKM with monomial potential
\cite{FKN,DVV,Mik,KMMMP} and turn into even more sophisticated
relations for a generic $V(x)$, especially when $s$-dependence is
also taken into account. For fixed $V(x)$ the different branches in
Kontsevich phase possess loop expansions and are further associated
with shifts $t_k \rightarrow T_k + t_k$, so that expansions are in
positive powers of $t$-variables with $T$ appearing in denominators
-- just like in the case of the Hermitian matrix model. The
$(p,q)$-model is the GKM with $V_p(x) = \frac{x^{p+1}}{p+1}$ in the
phase with $T_k\neq 0$ for $k=1, p+q$ only \cite{FKN1}.

\bigskip

$\bullet$ One of the most remarkable properties of GKM is the {\it
p-q duality} \cite{FKN,quas,KM}: the relation between partition
functions $Z_{p,q}$ and $Z_{q,p}$. It is not a literal coincidence
between the two branches of the partition function, instead they are
associated with two different coverings of one and the same spectral
curve and should coincide after an appropriate change of
time-variables.

\bigskip

$\bullet$ Since partition functions of all models, associated with
the Itzykson-Zuber integral (\ref{IZ}) possess the determinant
representations (\ref{IZev}), it is natural that they are
$\tau$-functions of the KP and Toda families in $t$-variables
\cite{GKM,versus,intmamo,mamo_r}. They also possess certain
integrability properties w.r.t. the $s$-variables
\cite{quas,versus,GKM2,Kr}. Thus, the entire theory of
Itzykson-Zuber-Kontsevich models is indeed a piece of theory of
stringy $\tau$-functions. These $\tau$-functions are, in fact,
closely related to the Hermitian model $\tau$-functions: both
classes belong to the same matrix model $M$-theory \cite{MthMamo}.

\bigskip

$\bullet$ The main questions to be addressed in the course of study of every
particular phase of every particular model are listed in the
following table:

\bigskip

\centerline{
\begin{tabular}{ccccc}
&& IZK integral $=$  && \\
&& partition function
$I = \exp \left(\sum_p g^{2p-2}F^{(p)}\right)$ && \\
&& $\downarrow$ && \\
&& Ward identities in the form && \\
&& of Virasoro and $W$ constraints
$\hat L(z) I = 0$&& \\
& $\swarrow$ && $\searrow$ & \\
genus-zero part $F^{(0)}$ of partition function \hspace{-2.0cm} &&&&
\hspace{-4.0cm}
multi-resolvents $\rho^{(\!p\,|m)}(z_1,\ldots,z_m)$ \\
$\downarrow$ &&& $\downarrow$ &
 \\
bare spectral curve $\Sigma_0$ && $\longrightarrow$
 & \hspace{-2.0cm} poly-differentials on $\Sigma_0$  &
 $\downarrow$
 \\
$\downarrow$ &&& $\downarrow$ &
\\
full spectral curve $\Sigma$, &&&& \hspace{-4.0cm}
Laplace transforms $\eta^{(p|m)}(x_1,\ldots,x_m)$ \\
describing the matrix-model $\tau$-function \hspace{-2.0cm} &&&&
\end{tabular}
}

\bigskip

The bare spectral curve $\Sigma_0$ is an important characteristic of
the branch of the partition function: different phases of the same
model differ by the shape of $\Sigma_0$. In fact, in order to
describe higher multi-resolvents in generic phases \cite{amm1,amm2},
the spectral curve should be made dependent on the genus expansion
parameter $g$, thus breaking the simple association between
$\Sigma_0$ and $F^{(0)}$. The study of this phenomenon can be one of
the clues to the construction of the last vertical arrow in the left
column, relating $\Sigma_0$ with the full spectral curve $\Sigma$.
As every KP/Toda $\tau$-function, the partition function is formally
associated with a point of the Grassmannian \cite{Gra} and, thus,
formally with some infinite genus Riemann surface: this is exactly
what we call $\Sigma$. The horizontal line in the center of the
table is a functorial map from complex curves to a hierarchical
family of poly-differentials, which can be described and studied
independently of other parts of the table. A big step in this
direction is described in \cite{EyKo}, but representation in terms
of free fields on $\Sigma_0$ is still lacking, even in the simplest
phases. Moreover, this map depends on additional conditions imposed
on the poly-differentials, which are actually related to the choice
between different branches of partition function, made also beyond
the genus zero approximation. Also lacking is a description of the
vertical arrow from poly-differentials to their Laplace transforms,
which should be very interesting, because the Laplace transforms
possess a very simple $m$-fold integral representation, at least, in
some phases \cite{ok}.

\section{Kontsevich model}

\subsection{Solving loop equations}
\setcounter{equation}{0}

\subsubsection{Kontsevich model: definitions}

We start with defining the Kontsevich model. As explained in the
Introduction, we define any matrix model partition function as a
solution to an infinite set of equations. In particular, the
Kontsevich partition function is defined to satisfy the continuous
Virasoro constraints: \be\label{K}
\hat L_-(z) Z(t) = 0, \\
\hat L_-(z) =
\left(:\hat J^2(z):\right)_-=\sum_n {\hat L_n(dz)^2\over z^{n+2}}=
\\= \frac{g^2}{8}\sum_{n=-1}^{+\infty}
\frac{(dz)^2}{z^{n+2}}\left( \sum_{k>0}
\left(2k+1\right)t_{2k+1}\frac{\p}{\p t_{2(k+n)+1}} + \frac{g^2}{2}\sum_{a+b=n-1}\frac{\p^2}{\p t_{2a+1}\p t_{2b+1}}+
\frac{\delta_{n,0}}{8}+\frac{\delta_{n,-1}t_1^2}{2g^2}
\right)\\
{\hat J}(z|t) = \frac{1}{4}
\sum_{k=0}^\infty \left\{\Big(2k+1\Big)t_{2k+1} z^{k-1/2}dz +
g^2\frac{dz}{z^{k+3/2}}\frac{\partial}{\partial t_{2k+1}}\right\}
\ee

In order to define the branch of the partition function, we shift
the times, \be t_{2k+1}=\tau_{2k+1}+T_{2k+1}\;\;\;0\le k\le N, \ee
and consider the partition function to be a formal power series in
the shifted times $\tau_{2k+1}$ that satisfies (\ref{K}).

Now we shall follow the line of paper \cite{amm1} and rewrite
(\ref{K}) in the form of loop equations that admit recursion
solving. To this end, we introduce
\begin{itemize}
\item
the loop operator \be \D(z)=\sum_{n\from
0}\frac{1}{z^{n+3/2}}\frac{\d}{\d \tau_{2n+1}} \ee
\item
the generating function for $T_k$ (a polynomial of degree $N$) \be
W(z)=\sum_{k=0}^N(2k+1)z^{k-1/2}T_{2k+1} \ee
\item the generating function for $\tau_k$ (a power
series)\be\label{u}
v(z)=\sum_{k=0}^{\infty}(2k+1)z^{k-1/2}\tau_{2k+1} \ee
\item the projector onto the negative part of series \be
P^-_z\left\{\sum_{k=-\infty}^{+\infty}z^ka_k\right\}=\sum_{k\le
-1}z^{k}a_k\hspace{25mm}P^+_z=1-P^-_z\ee
\item
the free energy and its topological expansion w.r.t. to the genus
$p$ \be Z(\tau)=e^{\frac{1}{g^2}\F(\tau)}\hspace{15mm}
\F=\sum_{p\from 0}{g^{2p}\F^{(p)}}\hspace{15mm}\F|_{\tau=0}=F[T]\ee
\item
the generating resolvent \be\label{G} G(z|\tau)=\D(z)\F(\tau)\ee
\item
and the multi-resolvents \be
\rho^{|m)}(z_1,\ldots,z_m)=\D(z_1)\cdots\D(z_m)\F|_{\tau=0} \ee \be
\rho^{(p|m)}(z_1,\ldots,z_m)=\D(z_1)\cdots\D(z_m)\F^{(p)}|_{\tau=0}
\ee
\item the set of $f$-functions generated by the $R$-check operator \be
 P^-_z(\aW(z)G(z|\tau))=\aW(z)G(z|\tau)-f(z|\tau)
\ee \be
 f(z|\tau)=P^+_z(\aW(z)G(z|\tau))=\Rc(z)\F(t)
\ee \be\label{Rcheck}
 \Rc(z)=\sum_{m=0}^{N-2}\sum_{k=m+2}^Nz^{k-m-2}(2k+1)T_{2k+1}\frac{\d}{\d
T_{2m+1}} \ee \be
f^{(p|m+1)}(z|z_1,\ldots,z_m)=\D(z_1)\cdots\D(z_m)\Rc(z)\F^{(p)}(\tau)|_{\tau=0}=\Rc(z)\rho^{(p|m)}(z_1,\ldots,z_m)
\ee
\end{itemize}

\bigskip

\subsubsection{The loop equation and recursion relations on the
multi-resolvents}

Now rewrite the Virasoro constraints (\ref{K}) in the form of the
loop equation for the resolvent \be
P^-_z(v(z)G(z))+\aW(z)G(z)-f(z)+\frac{1}{2}G^2(z)+\frac{g^2}{2}\D(z)G(z)+\frac{g^2}{8z^2}+\frac{(\tau_1+T_1)^2}{2z}=0
\label{loop} \ee Applying the operator $\D$ to this equation $k$
times, using the identity
 \be \D(x)P^-_z \left\{v(z)h(z)\right\}=2\d_x\left\{\frac{\left(\frac{x}{z}\right)^{\frac{1}{2}}
h(z)-h(x)}{z-x}\right\}+ P^-_z \left\{v(z)\D(x)h(z)\right\}\ee and
ultimately putting all $\tau_k=0$, one comes to the set of recursion
relations for the multi-resolvents,
\[
2\sum_{i=1}^k\d_{z_i}\left\{\frac{\left(\frac{z_i}{z}\right)^{\frac{1}{2}}\rho^{|k)}(z,z_1,\ldots,\hat{z_i},\ldots,z_k)-\rho^{|k)}(z_1,\ldots,z_k)}{z-z_i}\right\}-f^{|k+1)}(z|z_1,\ldots,z_k)
+\]
\[+\aW(z)\rho^{|k+1)}(z,z_1,\ldots,z_k)+\frac{1}{2}\sum_{k_1+k_2=k}\rho^{|k_1+1)}(z,z_{i_1},\ldots,z_{i_{k_1}})\rho^{|k_2+1)}(z,z_{j_1},\ldots,z_{j_{k_2}})+
\]
\be\label{rr1}
+\frac{g^2}{2}\rho^{|k+2)}(z,z,z_1,\ldots,z_k)+\frac{g^2}{8z}\delta_{k,0}+\frac{1}{z\prod_{i=1}^kz_i^{3/2}}\frac{1}{(2-k)!}T_1^{2-k}=0
\ee These recursive relations are invariant with respect to two
different scaling transformations with the following scaling
exponents: \be\label{grad1} \deg\tau_{2n+1}=n-1\hspace{2cm}\deg
g^2=\deg F=-3 \hspace{2cm} \deg T_{2n+1} =n-1\\ \deg
z=-1\hspace{1cm}\deg \nabla=\frac{5}{2}\hspace{1cm}\deg
\rho^{(p|k)}=-3+\frac{5}{2}k+3p
\ee and \be\label{grad2} \deg'g=\deg't_i=1\hspace{2cm}\deg'F=2 \\
\deg' z=0\hspace{1cm}\deg' \nabla=-1
\hspace{1cm}\deg'\rho^{(p|k)}=2-k-2p \ee

Making the genus expansion of the recursive equations (\ref{rr1}),
one obtains for the $g^{2p}$-term
\[
2\sum_{i=1}^k\dd_{z_i}\left\{\frac{\left(\frac{z_i}{z}\right)^{\frac{1}{2}}\rho^{(p|k)}(z,z_1,\ldots,\hat{z_i},\ldots,z_k)-\rho^{(p|k)}(z_1,\ldots,z_k)}{z-z_i}\right\}-f^{(p|k+1)}(z|z_1,\ldots,z_k)
+\]
\[+\aW(z)\rho^{(p|k+1)}(z,z_1,\ldots,z_k)+\frac{1}{2}\sum_{q=0}^p\sum_{{k_1+k_2=k
}}\rho^{(q|k_1+1)}(z,z_{i_1},\ldots,z_{i_{k_1}})\rho^{(p-q|k_2+1)}(z,z_{j_1},\ldots,z_{j_{k_2}})+
\]
\be
+\frac{1}{2}\rho^{(p-1|k+2)}(z,z,z_1,\ldots,z_k)+\frac{1}{8z}\delta_{k,0}\delta_{p,1}+\frac{1}{\prod_{i=1}^kz_i^{3/2}}\frac{1}{(2-k)!}T_1^{2-k}\delta_{p,0}=0
\label{rec_rel}\ee These double recursion relations (in $p$ and $k$)
can be used to determine all the multi-resolvents $\rho^{(p|k+1)}$
recursively. E.g., for $\rho^{(0|1)}$ we have a quadratic equation
\be
(\rho^{(0|1)}(z))^2+2W(z)\cdot\rho^{(0|1)}(z)-2\Rc(z)F^{(0)}[T]+T_1^2/z
\ee its solution being\footnote{Note that the non-meromorphic term
$W(z)dz$ in $\rho^{(0|1)}dz$ just cancels the singular part of the
$z$-expansion of $ydz$ at infinity.} \be
\rho^{(0|1)}(z)={-\aW(z)+y(z)}\ee where $y(z)$ is a multi-valued
function of $z$ \be y^2=(\aW(z))^2-\frac{T_1^2}{z}+2\Rc(z)F^{(0)}[T]
\label{curve} \ee

Making further iterations, the multi-density $\rho^{(p|k+1)}$ enters
(\ref{rec_rel}) linearly with the factor $y(z)$, and one can make
iterations, e.g., using some computer algebra system in the
following order: $(0|1)\rightarrow (1|0)\rightarrow (0|2)\rightarrow
(1|1)\rightarrow (2|0) \rightarrow (0|3)\rightarrow (1|2)
\rightarrow \cdots$

Note that the recursion relations contain a lot of ambiguity encoded
in the functions $f^{(p|1)}(z)=\Rc(z)F^{(p)}[T]$. Indeed, $F[T]$ can
be an arbitrary function that satisfies the two constraints $\check L_{-1}$, $\check L_0$
(these
are ${L}_{-1}$- and ${L}_0$-constraints with all $\tau_k=0$): \be
\sum_{k=1}^N(2k+1)T_{2k+1}\frac{\d F}{\d
T_{2k-1}}+\frac{T_1^2}{2}=0\hspace{2cm}\sum_{k=0}^N(2k+1)T_{2k+1}\frac{\d
F}{\d T_{2k+1}}+\frac{g^2}{8}=0 \label{init_constr} \ee Therefore,
the space of solutions to the loop equations (Virasoro constraints)
is parameterized by such functions $F[T]$.

\subsection{Solving the reduced Virasoro constraints}\label{sect_RVC}

The general solution of the second equation of (\ref{init_constr}),
i.e. of the $L_0$-constraint, is \be
F[T]=-\frac{g^2}{8(2N+1)}\log{T_{2N+1}}+\tilde{F}(\chi_1,\ldots,\chi_{N})
\label{red_L0_sol} \ee where \be
\chi_k=\frac{(2N-2k+1)!!}{(2N+1)!!}\cdot\frac{-2T_{2N-2k+1}}{(-2T_{2N+1})^{\frac{2N-2k+1}{2N+1}}}
\label{red_chi_def}\ee and $\tilde{F}$ is an arbitrary function.
Then the ${L}_{-1}$-constraint reads as \be\label{31}
\sum_{k=0}^{N-1}\chi_k\frac{\d\tilde{F}}{\d\chi_{k+1}}=-\frac{[(2N+1)!!]^2}{8}\chi_N^2
\ee where $\chi_0= 1$ is not an independent variable. Its general
solution is \be
\tilde{F}=\frac{[(2N+1)!!]^2}{8}\int_0^{\eta_{1}}{\chi_N^2(\tilde\eta_1,\eta_2\ldots,{\eta}_{N})
d\tilde{\eta}_{1}}+ \tilde{\tilde F}(\eta_2,\ldots,\eta_{N})
\label{red_L1_sol} \ee where $\tilde{\tilde F}$ is a new arbitrary
function, and we made the triangle change of
variables\footnote{Since this change of variables is triangle, the
variables $\chi_*$ can be equally well expressed through $\eta_*$.}
generated by the following relations between the generating
functions \be \chi (z)\equiv\sum_{k=1}^N \chi_k z^k\ \ \ \ \ \ \
\eta(z)\equiv\sum_{k=1}^N\eta_k z^k=\sum_{p=1}^\infty {(-)^{p}\over
p}\chi^p(z) \ee In other words, $\eta_k$ for $k<N+1$ are defined
from the equation $e^{\eta(z)}={1\over 1+\chi (z)}+O(z^{N+1})$ with
all $\chi_k$ and $\eta_k$ equal to zero if $k>N$.

In order to prove that (\ref{red_L1_sol}), indeed, solves
(\ref{31}), one suffices to note that, since
$$\eta_i=\oint_0\sum_{p=1}{(-)^p\over p}{\chi^p(z)\over z^{i+1}}dz$$
the $L_{-1}$-constraint (\ref{31}) reads as \be
\sum_{k=0}^{N-1}\chi_k\frac{\d\tilde{F}}{\d\chi_{k+1}}=
\sum_{k=0}^{N-1}\sum_i\chi_k{\d\eta_i\over\d\chi_{k+1}}\frac{\d\tilde{F}}{\d\eta_i}=
\sum_i\sum_{k=0}^{N-1}\sum_{p=1}(-)^{p}\chi_k\frac{\d\tilde{F}}{\d\eta_i}\oint_0{\d\chi
(z)\over\d\chi_{k+1}} {\chi^{p-1}(z)\over z^{i+1}}dz=\\=
\sum_i\frac{\d\tilde{F}}{\d\eta_i}\sum_{p=1}(-)^{p}\oint_0{\chi^{p-1}
(z)[1+\chi (z)]\over
z^{i}}dz=-\sum_i\frac{\d\tilde{F}}{\d\eta_i}\oint_0{1\over
z^{i}}dz=-\frac{\d\tilde{F}}{\d\eta_1}\label{34}\ee

Note that this choice of $\eta$-variables is in no way unique: e.g.,
one can equally well transform $\eta_2,\ldots,\eta_N$ to any new
$N-2$ variables without changing the formulas above. For instance,
one can request that the transformation is {\it linear} in all
$\chi_l$ with $l\ge 3$, \be \bar\eta_k=Q_k(\chi_1,\chi_2)+\sum_{l\ge
3}^k\chi_lP_{k,l}(\chi_1,\chi_2) \label{35}\ee and check that there exist
polynomials $Q_k$ and $P_{k,l}$ that preserve relation
(\ref{34}). Inserting (\ref{35}) into (\ref{34}), one immediately
obtains that these polynomials satisfy only the equations \be {\d
Q_k\over\d\xi_1}=-\chi_2 P_{k,3}\ \ \ \ \ \ {\d
P_{k,l}\over\d\xi_1}=-P_{k,l+1} \label{36}\ee where we changed
$\chi_{1,2}$ for new variables $\xi_{1,2}$ \be \xi_1\equiv\chi_1\ \
\ \ \ \xi_2\equiv\chi_2-{\chi_1^2\over 2} \ee Equations (\ref{36})
also have a lot of solutions. In particular, one can add to $Q_k$ an
arbitrary function of $\xi_2$. Choosing, e.g., this function to be
zero and $P_{k,k}=1$, one immediately obtains \be
P_{k,k-l}=(-)^l{\xi_1^l\over l!}\ \ \ \ \ \ \ \
Q_k=(-)^k\left[{\xi_1^{k-2}\over k-2}\xi_2+{\xi_1^k\over
2k(k-3)!}\right]\ee

\subsection{Example of transition: ${N=2 \rightarrow N=1}$
}\label{ell_rat}

Let us consider the limit of $T_5\to 0$ and see how the space of
solutions to the loops equations, which is parameterized by a
function of one variable in the case of $N=2$, reduces to the only
solution in the case of $N=1$.

For $N=2$ one has \be \chi_1=\frac{1}{5}\frac{-2T_3}{(-2T_5)^{3/5}}
\hspace{3cm}\chi_2= \frac{1}{15}\frac{-2T_1}{(-2T_5)^{1/5}}\ee \be
\eta_1=-\chi_1\hspace{3cm}\eta_2=\frac{1}{2}\chi_1^2-\chi_2 \ee \be
F[T]=-\frac{g^2}{40}\log
T_5-15^2\left(\frac{1}{60}\chi_1^5-\frac{1}{12}\chi_1^3\chi_2+\frac{1}{8}\chi_1\chi_2^2\right)+
\tilde{\tilde F}(\eta_2) \label{red_sol_N2} \ee In order to have a
smooth transition as $T_5$ goes to zero, one has to cancel the
singularity, i.e. $\tilde{\tilde F}(\eta_2)$ must have an
asymptotics \be \tilde{\tilde
F}(\eta_2)=15\sqrt{2}\cdot\eta_2^{5/2}-\frac{g^2}{48}\log\eta_2+
\mathcal{O}(1),\;\;\;\eta_2\rightarrow\infty \label{asympt_eta_0}\ee
and, therefore, one obtains \be F[T]^{(N=1)}=\lim_{T_5\rightarrow
0}F[T]^{(N=2)}=-\frac{g^2}{24}\log
T_1-\frac{1}{18}\frac{T_1^3}{T_3}+\const\ee which coincides with
(\ref{F_T_sphere}) below.

One may use the scaling symmetry (\ref{grad1}) in order to further
restrict the function $\tilde{\tilde F}$. Indeed, using that the
scaling dimension of $\eta_2$ is $-6/5$, one immediately comes to
the expansion \be \tilde{\tilde F}=\sum_{p=0}^\infty
g^{2p}C_p\eta_2^{5/2\cdot (1-p)}+C'_1g^2\log\eta_2
\label{elliptic_g_exp}\ee  {\bf if} assuming that the symmetry
(\ref{grad1}) does not change the solution, or, putting this
differently, that $\tilde{\tilde F}$ does not contain any additional
dimensional parameters changing under this symmetry transformation.
$C_0$ and $C_1'$ can be obtained by comparing (\ref{elliptic_g_exp})
with (\ref{asympt_eta_0}). Therefore, the asymptotic
(\ref{asympt_eta_0}) corresponds actually to the semiclassical limit
of $g\rightarrow 0$.

Geometrically this transition corresponds to degeneration of the
torus into the sphere. Indeed, formula (\ref{curve}) describes the
torus in the case of $N=2$. As it follows from (\ref{red_sol_N2})
and (\ref{asympt_eta_0}), $2\Rc(z)F^{(0)}[T]=10T_5\partial
F^{(0)}[T]/\partial T_1\to 0$ as $T_5\to 0$. Therefore,
(\ref{curve}) transforms under this transition exactly into the
sphere corresponding to $N=1$.

\subsection{Resolvents} \label{Resolvents_section}

As as explained above, starting from the one-point resolvent  \be
\rho^{(0|1)}(z)={-\aW(z)+y(z)} \ee one recursively calculates
further resolvents: \be
\rho^{(0|2)}(x_1,x_2)=\frac{1}{y(x_1)}\left(2\d_{x_2}\left[\frac{\left(\frac{x_2}{x_1}\right)^{\frac{1}{2}}\rho^{(0|1)}(x_1)-\rho^{(0|1)}(x_2)}{x_1-x_2}\right]-f^{(0|2)}(x_1|x_2)
+\frac{T_1}{x_1x_2^{3/2}}\right) \ee \be
\rho^{(1|1)}(x)=\frac{1}{y(x)}\left(\frac{1}{2}\rho^{(0|2)}(x,x)-f^{(1|1)}(x)
+\frac{1}{8x}\right) \ee etc.

These resolvents can be given a geometric meaning. Indeed, the
r.h.s. of (\ref{curve}) is actually a polynomial of degree $2N-1$.
Thus, equation (\ref{curve}) defines a hyperelliptic curve
$\Sigma_0=\mathcal{C}$ of genus $N-1$ in a generic case. (This genus
should not be confused with the genus corresponding to the expansion
of the free energy in powers of $g$, usually labeled by $p$.) This
bare spectral curve is actually essential for constructing all the
multi-resolvents because these are meromorphic multi-differentials
$\rho^{(p|m)}\equiv \rho^{(p|m)}(z_1,\cdots,z_m)dz_1\cdots dz_m$ on
this curve with specified singularities ($\rho^{(0|1)}$ and
$\rho^{(0|2)}$ are distinguished differentials playing a specific
role). Typically this leaves some room for adding holomorphic
differentials which exactly corresponds to the ambiguity in
solutions to the loop equations.

Indeed, because of equations (\ref{init_constr}), one has an
arbitrary function $F$ of $(N+1)-2=N-1$ variables, and, at each step
of recursive computation of the multi-resolvents, one finds in
$\rho^{(p|m)}$ some ambiguous terms $\sim \frac{\d^m F^{(p)}}{\d
T_{2n_1+1}\cdots \d T_{2n_m+1}},\;0\le n_i\le N-2$. Fixing these terms is
equivalent to fixing the periods of $\rho^{(p|m)}$. All these terms
are certainly fully determined by ${\d F^{(p)}\over\d T_{2n+1}}$, i.e. by
fixing the periods of $\rho^{(p|1)}(z)$.

As an example, consider the two-point resolvent. It can be rewritten
in the form
\begin{equation}
\rho^{(0|2)}=\rho_{hol}^{(0|2)}+\rho_{glob}^{(0|2)}-\rho_{loc}^{(0|2)}
\end{equation}
where $\rho^{(0|2)}_{hol}$ is a holomorphic bi-differential on the
curve $\mathcal{C}$, \be
\rho_{hol}^{(0|2)}(x_1,x_2)=\frac{\check{K}(x_1,x_2)F^{(0)}[T]\,dx_1dx_2}{y(x_1)y(x_2)}
\ee $\rho^{(0|2)}_{glob}$ is a meromorphic bi-differential on the
curve $\mathcal{C}$ that has the singularity at $x_1=x_2$ at the
{\it both} sheets of the curve of the following type: \be
\rho_{glob}^{(0|2)}\sim\frac{2dx_1dx_2}{(x_1-x_2)^2}+O(1)\label{rho_0_2_sing}
\ee At last,
$\rho^{(0|2)}_{loc}\equiv\left.\rho^{(0|2)}_{glob}\right|_{y(z)=\sqrt{z}}$
 has the same behaviour as
$\rho^{(0|2)}_{glob}$ at infinity and cancels the singularity
(\ref{rho_0_2_sing}) of $\rho^{(0|2)}_{glob}$ at one of the sheets
of the curve.

Manifestly, \be\label{rholocal}
\rho_{loc}^{(0|2)}=\frac{\left[x_1+x_2\right]dx_1dx_2}{\,x_1^{1/2}\,x_2^{1/2}\,(x_1-x_2)^2}
\ee \be
\rho_{glob}^{(0|2)}(x_1,x_2)=\frac{1}{y(x_1)y(x_2)(x_1-x_2)^2}((x_1+x_2)B(x_1,x_2)+(x_1-x_2)^2C(x_1,x_2)+\check{A}(x_1,x_2)F^{(0)}[T])
\ee \ni where \ba
B(x_1,x_2)=\frac{\aW(x_1)\aW(x_2)}{\sqrt{x_1x_2}}+\left[\frac{x_1}{x_2}+\frac{x_2}{x_1}-3\right]\frac{T_1^2}{x_1x_2}\\
C(x_1,x_2)=\left(T_1\left[\frac{\aW(x_2)}{\sqrt{x_2}}+\frac{\aW(x_1)}{\sqrt{x_1}}\right]+3T_1T_3\right)/(x_1x_2)\\
\check{K}(x_1,x_2)=\sum_{m=0}^{N-2}\sum_{m'=0}^{N'-2}\sum_{k=m+2}^N\sum_{k'=m'+2}^Nx_1^{k-m-2}x_2^{k'-m'-2}(2k+1)(2k'+1)T_{2k+1}T_{2k'+1}\frac{\d}{\d
T_{2m+1}}\frac{\d}{\d T_{2m'+1}}\\
\check{A}(x_1,x_2)=-\sum_{k=4}^N\sum_{n=0}^{k-4}\left\{x_1^{k-m-3}[3x_2-5x_1+2n(x_2-x_1)]+(x_2\leftrightarrow
x_1)\right\}(2k+1)T_{2k+1}\frac{\d}{\d
T_{2n+1}}+\nn \\
+4\left[5T_5\frac{\d}{\d T_1}+7T_7\frac{\d}{\d
T_1}+\frac{7}{2}(x_1+x_2)T_7\frac{\d}{\d T_1} \right] \ea  Note that
the numerator of $\rho^{(0|2)}_{glob}$ is actually a polynomial in
$x_1,\,x_2$.

Since $\rho^{(0|2)}_{hol}$ is a holomorphic bi-differential, the
second derivatives $\frac{\d^2 F[T]}{\d T_{2i+1}\d T_{2j+1}}$
(entering this differential) control the periods of $\rho^{(0|2)}$
and do not affect its singularities. Note that the number of
independent variables in $F[T]$ is equal to the genus of
$\mathcal{C}$: $N-1$.

Similarly one can deal with other resolvents in order to check that the
multi-resolvents $\rho^{(p|m)}=\rho^{(p|m)}(z_1,\cdots,z_m)dz_1\cdots dz_m$ are meromorphic
multi-differentials (except for the cases $(0|1)$ and $(0|2)$) on
the curve $\mathcal{C}$ and generically have poles of order
$6p+2m-4$ in points (and only in these point) where $y=0$. These
multi-resolvents can be further restricted with using symmetries
(\ref{grad1}) and (\ref{grad2}).

We discuss these general properties of multi-resolvents and its
applications in more details in the next section in the simplest
example of the Gaussian Kontsevich model.

\subsection{CFT representation} \label{CFT_section}

As already mentioned in the Introduction, the two-point function
$\rho^{(0|2)}_{glob}$ can be represented as a propagator in a
certain CFT:
\begin{equation}
 \rho_{glob}^{(0|2)}(z_1,z_2)=\langle \d X(z_1)\d X(z_2)\rangle \label{rho_0_2_CFT}
\end{equation}
where $X$ is some local field defined on $\mathbb{CP}^1$
parameterized by $z$. This is because there is a singularity at
$z_1=z_2$, i.e. when the arguments of the fields coincide. The CFT
is defined by the covering
$\mathcal{C}\stackrel{\pi}{\longrightarrow}\mathbb{CP}^1$. One way
is to consider the scalar field living on the $\mathcal{C}$ as a
collection of two fields $X_1$, $X_2$ living on the corresponding
sheets of the covering:
$\mathcal{C}\stackrel{(X_1,X_2)}{\longrightarrow}\mathbb{C}$ . Then
there will be a monodromy $X_1\leftrightarrow X_2$ when $z$ goes
around a branch point $y=0$. Then the field $X$ is a linear
combination of fields $X_1,\,X_2$ that diagonalizes this monodromy:
$X=X_1-X_2,\;X\leftrightarrow -X$.

To put this differently, let $z$ parameterize the whole world-sheet
now (i.e. topologically it would be a sphere), but the target space
is now an orbifold:
$\mathcal{C}/\mathbb{Z}_2\simeq\mathbb{CP}^1\stackrel{X}{\longrightarrow}\mathbb{C}/\mathbb{Z}_2$.
The branching points $y=0$ are now just points where string wraps
around the $\mathbb{Z}_2$-fixed point.

Both of these approaches can be actually described in the same way
via the branching point operators of \cite{Zam,CFT}. To this end,
one needs to introduce the twist field $\sigma_{1/2}(w,\bar{w})$
with the following operator product expansion (OPE):
\begin{equation}
 \d X(z)\sigma_{1/2}(w,\bar{w})\sim (z-w)^{-1/2}\tau_{1/2}(w,\bar{w})+\ldots
\end{equation}
where $\tau_{1/2}$ is sometimes called excited twist field. Then,
one can write
\begin{equation}
 \langle \d X(z_1)\d X(z_2)\rangle=\left\langle \d X(z_1)\d X(z_2)\prod_{y(w_i)=0}\sigma_{1/2}(w_i,\bar{w}_i)\right\rangle_0
\label{rho_0_2_CFT_branch}
\end{equation}
By $\left\langle\cdot\right\rangle_0$ we denote the correlator in
the ordinary CFT on $\mathbb{CP}^1$. This would provide us with the
necessary structure of singularities
\begin{equation}
 \langle \d X(z_1)\d X(z_2)\rangle\sim\frac{2}{(z_1-z_2)^2}+O(1),\;\;\;z_1\rightarrow z_2
\end{equation}
\begin{equation}
 \langle \d X(z_1)\d X(z_2)\rangle\sim (z_i-w_j)^{-1/2},\;\;\;z_i\rightarrow w_j,\;\;\;y(w_j)=0
\end{equation}
from which one can deduce (\ref{rho_0_2_CFT}).

One can also include $\rho^{(0|2)}_{hol}$ in this correlator. It
would control the global monodromy properties of the filed $X$, i.e.
how it changes when $z$ goes around the cycles on $\mathcal{C}$.

At last,
\begin{equation}
  \rho_{loc}^{(0|2)}(z_1,z_2)=\left\langle \d X(z_1)\d X(z_2)\sigma_{1/2}(0,0)\right\rangle_0
\end{equation}
since $\rho_{loc}^{(0|2)}(z_1,z_2)$ knows nothing about the
branching points $w_i$, see (\ref{rholocal}).

\section{Gaussian branch of Kontsevich model}
\setcounter{equation}{0}

\subsection{Specific of the Gaussian branch}

In this section we consider the special, simplest case with only the
first two times non-perturbatively turned on ($N=1$). Then, the
$R$-check operator (\ref{Rcheck}) identically vanishes \be
\Rc(z)=0\Rightarrow f^{(p|k)}=0,\;\;\forall p,k \ee Therefore, there
are no ambiguities in resolvents in this case. This key feature
suggests a separate study of this distinguished case. This case is
also a counterpart of the Gaussian branch of the Hermitian matrix
model, hence, we call it the Gaussian branch. To avoid
misunderstanding, note that it has nothing to do with the Gaussian
integral!

Given $T_1=a$ and $T_3=-\frac{1}{3}M$, the solution of
(\ref{init_constr}) is \be
F[T]=F(M,a)=\frac{1}{6}\frac{a^3}{M}-\frac{g^2}{24}\log\frac{M}{M_0}
\label{F_T_sphere}\ee \be
Z(M,a)|_{\tau=0}=\left(\frac{M}{M_0}\right)^{-\frac{1}{24}}e^{\frac{1}{6g^2}\frac{a^3}{M}}\;\;\;\;\;(Z|_{t_{2k+1}=-\frac{1}{3}M_0\delta_{k,1}}=1)
\ee and the curve is \be \hspace{15mm}y^2=M^2(z-s)
\hspace{25mm}\left(s=\frac{2a}{M}\right)\ee \be
\aW(z)=\frac{a}{\sqrt{z}}-M\sqrt{z}=\frac{M(z-s/2)}{\sqrt{z}} \ee
Therefore, the resolvents non-trivially depend only on one parameter
$s$ ($M$ can be effectively removed by rescalings). To simplify
formulas, we consider from now on the redefined curve \be Y(z)\equiv
y(z)/M=\sqrt{z-s} \label{koncurve}\ee

In the Gaussian case, the recurrent relation (\ref{rec_rel}) can be
simplified. More concretely, for sufficiently large indices it can
be written in the form
\[y(z)\rho^{(p|k+1)}(z,z_1,\ldots,z_k)=-2\sum_{i=1}^k\d_{z_i}\frac{\rho^{(p|k)}(z_1,\ldots,z_k)}{z-z_i}+\]
\be +\frac{1}{2}\sum_{\hbox{\parbox{17mm}{\vspace{-4mm}\center \tiny
$k_1\!+\!k_2\!=\!k$\\$\!p_1\!+\!p_2\!=\!p$\\$k_i+p_i>0$
}}}\rho_{\tt}^{(p_1|k_1+1)}(z,z_{i_1},\ldots,z_{i_{k_1}})\rho_{\tt}^{(p_2|k_2+1)}(z,z_{j_1},\ldots,z_{j_{k_2}})+
\frac{1}{2}\rho^{(p-1|k+2)}(z,z,z_1,\ldots,z_k)\label{rec_rel_s}\ee
\ni where the subscript $(\cdot)_{\tt}$ means that one has to
replace $\rho^{(0|2)}$ with $\rho_{glob}^{(0|2)}$ leaving all other
$\rho$'s unchanged.

Indeed, for large enough indices (\ref{rec_rel}) contains, in the
Gaussian case, only four terms. The only term containing
$\rho^{(0|1)}$ combines with that containing $W(z)$ to produce
$y(z)$, while $\rho_{loc}^{(0|2)}$ in the sum quadratic in $\rho$'s
cancels the non-meromorphic
($\sim{\left(\frac{z_i}{z}\right)^{\frac{1}{2}}}$) part of the first
term.

Recurrent relations (\ref{rec_rel_s}) celebrate an important
property that leads to drastic simplifications in the Gaussian case,
which allows one to get rid of the only parameter $s$:

\paragraph{Important formula:}
{\it
All the $s$-dependence of the resolvents (except for $p=0$, $m=1$
and $p=0$, $m=2$ cases) is actually encoded only in the
 differences $z_i-s$:
\be\label{97}
\rho^{(p|m)}(z_1,\ldots,z_m|s)=\rho^{(p|m)}(z_1-s,\ldots,z_m-s|0)
\ee
}

\bigskip

In order to prove this formula, let us denote though
$l_{-1}$ the first-order part of the differential
operator $L_{-1}$:
$l_{-1}=\sum\limits_{k=1}^\infty(2k+1)t_{2k+1}\frac{\d}{\d
t_{2k-1}}$, $L_{-1}=l_{-1}+\frac{t_1^2}{2g^2}$. Then, the
$L_{-1}$-constraint on $\F$ is
\begin{equation}
 l_{-1}\F=-\frac{t_1^2}{2} \label{l_-1_on_F}
\end{equation}
To prove (\ref{97}), one suffices to note that
\begin{equation}
 [\,l_{-1},\D(z)]=2\frac{\d}{\d z}\D(z)\hspace{15mm}\hbox{and}\hspace{15mm}l_{-1}|_{\tau=0}=3\,T_3\frac{\d}{\d T_1}=-2\frac{\d}{\d s}
\end{equation}
and, using these formulae, to show immediately that $-\frac{\d}{\d
s}$ and $\sum\limits_i\frac{\d}{\d z_i}$ acting on
\begin{equation}
 \rho^{(p|m)}(z_1,\ldots,z_m)\eqdef \D(z_1)\cdots\D(z_m)\F^{(p)}|_{\tau=0}
\end{equation}
are equal to each other whenever $(p,m)\neq(0,1)\;\hbox{or}\;(0,2)$
(so that one can ignore the r.h.s. of (\ref{l_-1_on_F})).

Formula (\ref{97}) can be also proved by induction using the recursive
relations (\ref{rec_rel_s}). Indeed, the claim is correct for
$\rho_{\tt}^{(0|2)}$ by an immediate check. Further, if all the $\rho$'s
in the r.h.s. of (\ref{rec_rel_s}) enjoy the property (\ref{97}) (by
the induction assumption), this is also true for $z-z_i$ and $y(z)$
and, thus, for $\rho^{(p|k+1)}$ in the l.h.s.

We can use $Y$ as a standard coordinate on our $\mathbb{CP}^1$,
since $z=Y^2+s$. Then (\ref{97}) says that the densities can be
written in terms of $Y$'s only. Thus, the case of arbitrary $s$ is,
in a sense, equivalent to the $s=0$ case.

\subsection{Resolvents}

In the next subsection, we present for a reference manifest
expressions for several first densities. They all can be obtained
recursively using (\ref{rec_rel_s}), by hands or with the help of
computer (using, e.g., \textsc{maple}).

\subsubsection{First resolvents}

The one-point resolvents:

\begin{itemize}

\item Genus $p=0$  \be
\rho^{(0|1)}(z|s)=M\left(\sqrt{z}-\frac{s}{2\sqrt{z}}-Y(z)\right)
=M\frac{s}{2}\sum_{n=1}^{\infty}\frac{s^n}{z^{n+1/2}}\frac{\Gamma(n+1/2)}{(n+1)!\,\Gamma(1/2)}
\ee

\item Genus $p=1$ \be {\rho^{(1|1)}}(z|s)=
\frac{1}{8}\frac{1}{M}\frac{1}{Y^5(z)} \ee

\item Genus $p=2$ \be {\rho^{(2|1)}}(z|s)=
\frac{105}{128}\frac{1}{M^3}\frac{1}{Y^{11}(z)} \ee

\item Genus $p=3$ \be {\rho^{(3|1)}}(z|s)=
\frac{25025}{1024}\frac{1}{M^5}\frac{1}{Y^{17}(z)} \ee

\item Genus $p=4$ \be {\rho^{(4|1)}}(z|s)=
\frac{56581525}{32768}\frac{1}{M^7}\frac{1}{Y^{23}(z)} \ee

\end{itemize}

The list of resolvents grouped by the genus $p$ (we put $s=0$ in all
resolvents but $\rho^{(0|1)}$ and $\rho^{(0|2)}$; the $s$-dependence
can be easily restored using formula (\ref{97})):

\begin{itemize}
\item Genus $p=0$  \be \rho^{(0|1)}(z|s)=M\left(-\sqrt{z}+\frac{s}{2\sqrt{z}}-Y(z)\right)\ee

\be {\rho^{(0|2)}}({z_{1}}, \,{z_{2}}|s)=
\frac{z_1+z_2-2s}{(z_1-z_2)^2Y(z_1)Y(z_2)}-\frac{z_1+z_2}{z_1^{1/2}z_2^{1/2}(z_1-z_2)^2}
 \label{rho_0_2_g}\ee

\be \left(\rho^{(0|2)}(z,z|s)=\frac{(z-s/2)(s/2)}{Y^2(z)} \right)\ee

\be {\rho^{(0|3)}}({z_{1}}, \,{z_{2}}, \,{z_{3}}|s)={\displaystyle
\frac {1}{M}} \,{\displaystyle \frac {1}{Y^3(z_1)Y^3(z_2)Y^3(z_3) }}
\ee

\ba {\rho^{(0|4)}}({z_{1}}, \,{z_{2}}, \,{z_{3}}, \,{z_{4}}|s=0)=
{\displaystyle \frac {3}{M^2}}
\frac{1}{z_1^{5/2}z_2^{5/2}z_3^{5/2}z_4^{5/2}}\cdot({z_{1}}\,{z_{2}}\,{z_{3}}
+ {z_{1}}\,{z_{2} }\,{z_{4}} + {z_{1}}\,{z_{3}}\,{z_{4}}+
{z_{2}}\,{z_{3}}\,{z_{4}}) \ea

\ba
\rho^{(0|5)}(z_1,z_2,z_3,z_4,z_5|0)=\frac{3}{{M}^{3}\,{z_{{1}}}^{7/2}{z_{{2}}}^{7/2}{z_{{3}}}^{7/2}{z_{{4}}}^{7/2}{z_{{5}}}^{7/2}} \,(6\,{z_{{1}}}^{2}z_{{5}}{z_{{4}}}^{2}{z_{{3}}}^{2}z_{{2}}+5\,{z_{{1}}}^{2}{z_{{2}}}^{2}{z_{{5}}}^{2}{z_{{3}}}^{2}\\
\mbox{}+5\,{z_{{1}}}^{2}{z_{{2}}}^{2}{z_{{3}}}^{2}{z_{{4}}}^{2}+6\,{z_{{1}}}^{2}{z_{{2}}}^{2}{z_{{5}}}^{2}z_{{4}}z_{{3}}+6\,{z_{{1}}}^{2}z_{{2}}{z_{{5}}}^{2}{z_{{4}}}^{2}z_{{3}}+6\,{z_{{1}}}^{2}{z_{{5}}}^{2}z_{{4}}{z_{{3}}}^{2}z_{{2}}+5\,{z_{{1}}}^{2}{z_{{5}}}^{2}{z_{{4}}}^{2}{z_{{3}}}^{2}\\+6\,{z_{{1}}}^{2}{z_{{2}}}^{2}{z_{{3}}}^{2}z_{{4}}z_{{5}}+6\,{z_{{1}}}^{2}{z_{{2}}}^{2}z_{{5}}{z_{{4}}}^{2}z_{{3}}+5\,{z_{{1}}}^{2}{z_{{2}}}^{2}{z_{{5}}}^{2}{z_{{4}}}^{2}+6\,z_{{1}}{z_{{2}}}^{2}z_{{5}}{z_{{4}}}^{2}{z_{{3}}}^{2}+6\,z_{{1}}z_{{2}}{z_{{5}}}^{2}{z_{{4}}}^{2}{z_{{3}}}^{2}\\+6\,z_{{1}}{z_{{2}}}^{2}{z_{{5}}}^{2}z_{{4}}{z_{{3}}}^{2}+6\,z_{{1}}{z_{{5}}}^{2}{z_{{4}}}^{2}z_{{3}}{z_{{2}}}^{2}+5\,{z_{{2}}}^{2}{z_{{5}}}^{2}{z_{{3}}}^{2}{z_{{4}}}^{2})
\ea

 \item Genus $p=1$ \be {\rho^{(1|1)}}(z|s)=
\frac{1}{8}\frac{1}{M}\frac{1}{Y^5(z)} \ee

\be {\rho^{(1|2)}}({z_{1}}, \,{z_{2}}|0)={\displaystyle \frac
{1}{8M^2}} \, {\displaystyle \frac { 5\,{z_{1}}^{2} +
3\,{z_{1}}\,{z_{2}} + 5\,{z_{2}}^{2 }\,}{z_1^{7/2}z_2^{7/2}}} \ee

\ba {\rho^{(1|3)}}({z_{1}}, \,{z_{2}}, \,{z_{3}}|0)={\displaystyle
\frac {1}{8M^3}}
\frac{1}{z_1^{9/2}z_2^{9/2}z_3^{9/2}}\cdot(35\,{z_{1}}^{3}\,
{z_{2}}^{3} + 30\,{z_{1}}^{3}\,{z_{2}}^{2}\,{z_{3}} + 30\,{z_{1}}
^{3}\,{z_{2}}\,{z_{3}}^{2} + 35\,{z_{1}}^{3}\,{z_{3}}^{3} \\+ 30\,{
z_{1}}^{2}\,{z_{2}}^{3}\,{z_{3}} +
18\,{z_{1}}^{2}\,{z_{2}}^{2}\,{z_{3}}^{2} + 30\,{z_{1}}
^{2}\,{z_{2}}\,{z_{3}}^{3} + 30\,{z_{1}}\,{z_{2}}^{3}\,{z_{3}}^{2 }
+ 30\,{z_{1}}\,{z_{2}}^{2}\,{z_{3}}^{3} + 35\,{z_{2}}^{3}\,{z_{
3}}^{3}) \ea

\ba
\rho^{(1|4)}(z_1,z_2,z_3,z_4|0)=\frac{3}{8\,{M}^{4}\,{{z_{{1}}}^{11/2}{z_{{2}}}^{11/2}{z_{{3}}}^{11/2}{z_{{4}}}^{11/2}}} (105\,{z_{{1}}}^{4}{z_{{2}}}^{4}z_{{3}}{z_{{4}}}^{3}\\+100\,{z_{{1}}}^{4}{z_{{2}}}^{2}{z_{{3}}}^{4}{z_{{4}}}^{2}+105\,{z_{{1}}}^{4}{z_{{4}}}^{4}{z_{{3}}}^{4}+90\,{z_{{1}}}^{3}{z_{{2}}}^{2}{z_{{3}}}^{4}{z_{{4}}}^{3}+105\,{z_{{3}}}^{4}{z_{{2}}}^{3}z_{{4}}{z_{{1}}}^{4}+54\,{z_{{3}}}^{3}{z_{{2}}}^{3}{z_{{4}}}^{3}{z_{{1}}}^{3}\\+100\,{z_{{1}}}^{2}{z_{{2}}}^{2}{z_{{3}}}^{4}{z_{{4}}}^{4}+90\,{z_{{3}}}^{2}{z_{{2}}}^{3}{z_{{4}}}^{3}{z_{{1}}}^{4}+90\,{z_{{3}}}^{3}{z_{{2}}}^{3}{z_{{4}}}^{2}{z_{{1}}}^{4}+100\,{z_{{1}}}^{4}{z_{{2}}}^{2}{z_{{3}}}^{2}{z_{{4}}}^{4}+90\,{z_{{2}}}^{3}{z_{{3}}}^{2}{z_{{4}}}^{4}{z_{{1}}}^{3}\\+105\,{z_{{3}}}^{4}{z_{{2}}}^{4}z_{{4}}{z_{{1}}}^{3}+90\,{z_{{2}}}^{4}{z_{{3}}}^{3}{z_{{4}}}^{2}{z_{{1}}}^{3}+105\,{z_{{3}}}^{3}{z_{{2}}}^{4}z_{{4}}{z_{{1}}}^{4}+90\,{z_{{3}}}^{4}{z_{{2}}}^{3}{z_{{4}}}^{3}{z_{{1}}}^{2}+90\,{z_{{3}}}^{4}{z_{{2}}}^{3}{z_{{4}}}^{2}{z_{{1}}}^{3}\\
+90\,{z_{{1}}}^{3}{z_{{2}}}^{2}{z_{{3}}}^{3}{z_{{4}}}^{4}+105\,{z_{{1}}}^{4}{z_{{4}}}^{4}z_{{3}}{z_{{2}}}^{3}+90\,{z_{{1}}}^{4}{z_{{2}}}^{2}{z_{{3}}}^{3}{z_{{4}}}^{3}+90\,{z_{{2}}}^{4}{z_{{3}}}^{3}{z_{{4}}}^{3}{z_{{1}}}^{2}+105\,{z_{{2}}}^{4}{z_{{3}}}^{3}{z_{{4}}}^{4}z_{{1}}\\+90\,{z_{{2}}}^{4}{z_{{3}}}^{2}{z_{{4}}}^{3}{z_{{1}}}^{3}+100\,{z_{{2}}}^{4}{z_{{3}}}^{2}{z_{{4}}}^{4}{z_{{1}}}^{2}+100\,{z_{{2}}}^{4}{z_{{3}}}^{4}{z_{{4}}}^{2}{z_{{1}}}^{2}+100\,{z_{{2}}}^{4}{z_{{3}}}^{2}{z_{{4}}}^{2}{z_{{1}}}^{4}+105\,{z_{{1}}}^{3}{z_{{4}}}^{4}{z_{{3}}}^{4}z_{{2}}\\+105\,{z_{{1}}}^{4}{z_{{4}}}^{3}{z_{{3}}}^{4}z_{{2}}+105\,{z_{{2}}}^{4}{z_{{3}}}^{4}{z_{{4}}}^{3}z_{{1}}+105\,{z_{{2}}}^{4}{z_{{1}}}^{4}{z_{{3}}}^{4}+105\,{z_{{2}}}^{3}{z_{{3}}}^{4}{z_{{4}}}^{4}z_{{1}}+90\,{z_{{2}}}^{3}{z_{{3}}}^{3}{z_{{4}}}^{4}{z_{{1}}}^{2}\\+105\,{z_{{2}}}^{4}{z_{{3}}}^{4}{z_{{4}}}^{4}+105\,{z_{{1}}}^{4}{z_{{4}}}^{4}{z_{{3}}}^{3}z_{{2}}+105\,{z_{{2}}}^{4}{z_{{1}}}^{4}{z_{{4}}}^{4}+105\,{z_{{2}}}^{4}z_{{3}}{z_{{4}}}^{4}{z_{{1}}}^{3})
\ea

\item Genus $p=2$ \be {\rho^{(2|1)}}(z|s)=
\frac{105}{128}\frac{1}{M^3}\frac{1}{Y^{11}(z)} \ee

\ba {\rho^{(2|2)}}({z_{1}}, \,{z_{2}}|0)= {\displaystyle \frac
{35}{128M^4}} \frac{1}{z_1^{13/2}z_2^{13/2}}\cdot( 33\,{z_{1}}^{5}
 + 27\,{z_{1}}^{4}\,{z_{2}} \nn\\+ 29\,{z_{1}}^{3}\,{z_{2}}^{2}
 + 29\,{z_{1}}^{2}\,{z_{2}}^{3} + 27\,{z_{1}}\,{z_{2}}^{4} + 33\,
{z_{2}}^{5}) \ea

\ba
\rho^{(2|3)}(z_1,z_2,z_3|0)=\frac{35}{128\,{M}^{5}\,{z_{{1}}}^{15/2}{z_{{2}}}^{15/2}{z_{{3}}}^{15/2}}\, (396\,{z_{{1}}}^{5}z_{{2}}{z_{{3}}}^{6}\\
+396\,{z_{{2}}}^{6}{z_{{3}}}^{5}z_{{1}}+396\,{z_{{1}}}^{2}{z_{{2}}}^{6}{z_{{3}}}^{4}+406\,{z_{{2}}}^{6}{z_{{1}}}^{3}{z_{{3}}}^{3}+396\,{z_{{2}}}^{6}{z_{{1}}}^{4}{z_{{3}}}^{2}+396\,{z_{{2}}}^{6}{z_{{1}}}^{5}z_{{3}}\\+396\,{z_{{2}}}^{5}{z_{{1}}}^{6}z_{{3}}+348\,{z_{{1}}}^{4}{z_{{2}}}^{5}{z_{{3}}}^{3}+324\,{z_{{1}}}^{5}{z_{{2}}}^{5}{z_{{3}}}^{2}+396\,{z_{{3}}}^{6}{z_{{2}}}^{5}z_{{1}}+324\,{z_{{3}}}^{5}{z_{{2}}}^{5}{z_{{1}}}^{2}\\+348\,{z_{{3}}}^{4}{z_{{2}}}^{5}{z_{{1}}}^{3}+396\,{z_{{3}}}^{6}{z_{{2}}}^{4}{z_{{1}}}^{2}+429\,{z_{{2}}}^{6}{z_{{3}}}^{6}+348\,{z_{{3}}}^{5}{z_{{2}}}^{4}{z_{{1}}}^{3}+396\,{z_{{1}}}^{6}z_{{2}}{z_{{3}}}^{5}\\
+396\,{z_{{2}}}^{2}{z_{{1}}}^{4}{z_{{3}}}^{6}+324\,{z_{{2}}}^{2}{z_{{1}}}^{5}{z_{{3}}}^{5}+406\,{z_{{1}}}^{3}{z_{{2}}}^{3}{z_{{3}}}^{6}+348\,{z_{{1}}}^{4}{z_{{2}}}^{3}{z_{{3}}}^{5}+429\,{z_{{1}}}^{6}{z_{{3}}}^{6}\\+348\,{z_{{1}}}^{5}{z_{{3}}}^{4}{z_{{2}}}^{3}+396\,{z_{{2}}}^{2}{z_{{1}}}^{6}{z_{{3}}}^{4}+360\,{z_{{2}}}^{4}{z_{{1}}}^{4}{z_{{3}}}^{4}+406\,{z_{{1}}}^{6}{z_{{2}}}^{3}{z_{{3}}}^{3}\\
\mbox{}+396\,{z_{{2}}}^{4}{z_{{1}}}^{6}{z_{{3}}}^{2}+348\,{z_{{2}}}^{4}{z_{{1}}}^{5}{z_{{3}}}^{3}+429\,{z_{{1}}}^{6}{z_{{2}}}^{6})
\ea

\item Genus $p=3$ \be {\rho^{(3|1)}}(z|s)=
\frac{25025}{1024}\frac{1}{M^5}\frac{1}{Y^{17}(z)} \ee

\ba
\rho^{(3|2)}(z_1,z_2|0)=\frac{35}{1024\,{M}^{6}\,{z_{{1}}}^{19/2}{z_{{2}}}^{19/2}}
(12155\,{z_{{1}}}^{8}+10725\,{z_{{1}}}^{7}z_{{2}}+11011\,{z_{{2}}}^{2}{z_{{1}}}^{6}+11066\,{z_{{1}}}^{5}{z_{{2}}}^{3}\\+10926\,{z_{{1}}}^{4}{z_{{2}}}^{4}+11066\,{z_{{2}}}^{5}{z_{{1}}}^{3}+11011\,{z_{{1}}}^{2}{z_{{2}}}^{6}+10725\,{z_{{2}}}^{7}z_{{1}}+12155\,{z_{{2}}}^{8})
\ea

\end{itemize}

\subsubsection{Resolvents: general formulae and relations}
For generic $p$ and $m$, $(p,m)\neq(0,1)\;\hbox{or}\;(0,2)$, in the
Gaussian case all the resolvents are of the following form:
 \be {\rho^{(p|m)}}(z_1,\ldots,z_m|0)=
\frac{1}{M^{2p+m-2}}\left(\frac{Q_{p,m}(\{z_i\})}{\prod_{i=1}^mY^{6p+2m-3}(z_i)}\right)
\label{rho_p_m} \ee \ni where $Q_{p,m}$ are homogeneous symmetric
polynomials in $\{z_i\}$ of degree  $\deg Q_{p,m}=(m-1)(m+3p-3)$.

For $m=1$ $\deg Q_{p,m}=0$, and $Q_{p,m}$ is just a constant
 \be {\rho^{(p|1)}}(z|s)=
\frac{(6p-3)!!}{2^{3p}3^pp!}\frac{1}{M^{2p-1}}\frac{1}{Y^{6p-1}(z)}
\label{rho_p_1} \ee so that the formal power series
$\rho^{|1)}(z)=\sum_{p=0}^{\infty}g^{2p}\rho^{(p|1)}(z)$ can be
converted into the hypergeometric function. Its $z$-expansion is \be
{\rho^{(p|1)}}(z)=
\frac{(6p-3)!!}{2^{3p}3^pp!}\frac{1}{M^{2p-1}}\frac{1}{z^{3p-1/2}}\sum_{n=0}^{\infty}\frac{s^n}{z^n}\frac{\Gamma(n+3p-1/2)}{n!\,\Gamma(3p-1/2)}
\label{rho_p_1_expansion} \ee \ni which gives the general formula
for the $\langle(\sigma_0)^n\sigma_{3p-2+n}\rangle$ intersection
numbers \cite{Ko}.

There is also a general formula for the genus zero ($p=0$)
multi-resolvents: \be
 \rho^{(0|k)}(\{z_i\}_{i=1}^k)=\frac{1}{M^{k-2}}\,(2\d_s)^{k-3}\frac{1}{Y^3(z_1)\cdots Y^3(z_k)}
\label{rho_0_k_g} \ee A general formula for the $p=1$
multi-resolvents is less explicit. It can be written with the help
of the generating resolvent (\ref{G}) treated as a functional of
$\phi(z)$ such that $d\phi (z)/dz\equiv \frac{v(z)}{2}$, (\ref{u}):
\begin{equation}\label{rho_1_k_g_rough}
  G^{(0)}[\phi]=\sum_{k=0}^{\infty}\frac{1}{M^k\,k!}\prod_{i=1}^k\oint_Cdz_i\phi(z_i)\cdot
  \,(2\d_s)^k\frac{1}{\prod_{j=1}^kY^3(z_j)}=
\end{equation}
$$
  = \left.\sum_{k=0}^\infty \frac{(2\partial_s)^k}{M^kk!}
  \left(\oint_C
\frac{\phi(z)dz}{(z^2-s)^{3/2}}\right)^k \right|_{s=0} =
\left.\oint_0 e^{2t\partial_s} \frac{dt}{t - {1\over M}\oint_C
\frac{\phi(z)dz}{(z^2-s)^{3/2}}}\right|_{s=0} = \oint_0 \frac{dt}{t
- {1\over M}\oint_C \mu(z)\phi(z)dz}
$$
where $\mu(z) \equiv (z^2-2t)^{-3/2}$ and the contour $C$ encircles
$\infty$. Then, the logarithm of the generating resolvent generates
the $p=1$ multi-resolvents
\begin{equation}
\frac{1}{24}\log G^{(0)}[\phi] = \sum_{k=0}^\infty \frac{1}{k!}
\prod_{i=1}^k \left(\oint dz_i \phi(z_i)\right)
 \rho^{(1|k)}(z_1,\ldots,z_k),
\label{rho_1_k_g_connected}
\end{equation}
i.e. the genus one multi-resolvents are connected parts (up to the
factor of 24) of the $k$-point function generated by the generating
resolvent,
$$
\rho^{(1|k)}(z_1,\ldots,z_k) = \left.\frac{1}{24} \frac{\delta \log
G^{(0)}\{\phi\}} {\delta\phi(z_1)\ldots\delta\phi(z_k)}
\right|_{\phi = 0}
$$
For example,
$$
\rho^{(1|1)}(z) = \left. \frac{1}{24}\frac{\delta
G^{(0)}\{\phi\}}{\delta\phi(z)} \right|_{\phi = 0} =
\frac{1}{24M}\oint \frac{\mu(z)dt}{t^2} = \frac{1}{24M}\left.\dot
\mu(z)\right|_{t=0} =
\frac{1}{8Mz^{5/2}}
$$
$$
\rho^{(1|2)}(z_1,z_2) = \left. \frac{1}{24}\left(\frac{\delta^2
G^{(0)}\{\phi\}} {\delta\phi(z_1)\delta\phi(z_2)} - \frac{\delta
G^{(0)}\{\phi\}}{\delta\phi(z_1)} \frac{\delta
G^{(0)}\{\phi\}}{\delta\phi(z_2)}\right) \right|_{\phi = 0} =
$$
$$=
\frac{1}{24M^2}\left(2\oint \frac{\mu(z_1)\mu(z_2)dt}{t^3} -
\oint\frac{\mu(z_1)dt}{t^2} \oint\frac{\mu(z_2)dt}{t^2}\right) = $$
$$ =
\frac{1}{24M^2}\left.\left( \frac{2\big(\ddot \mu(z_1)\mu(z_2) +
2\dot \mu(z_1)\dot \mu(z_2) + \mu(z_1)\ddot \mu(z_2)\big)}{2} - \dot
\mu(z_1)\dot \mu(z_2) \right)\right|_{t=0} = \frac{5z_1^2 + 3z_1z_2
+ 5z_2^2}{8M^2 z_1^{7/2}z_2^{7/2}}
$$

\subsubsection{Proofs and comments} \label{gauss_proofs_and_comments}

In the forthcoming considerations, we use the symmetries
(\ref{grad1})\footnote{For $M=1,\;s=0$ this symmetry requirements
are equivalent to the standard claim that the free energy has an
expansion in $\tau$ such that \be F(\tau)=\sum
g^{2p}\tau_{2n_1+1}\cdots\tau_{2n_k+1}\cdot C^{(p)}_{n_1\cdots
n_k},\hspace{15mm}\sum_{i=1}^k(n_i-1)=3p-3 \ee \ni $C$'s being some
\textit{numerical} constants. This is usually derived \cite{Ko}
from the fact that the intersection number of a collection of forms
is non-zero only when the sum of their degrees is equal to the
dimension of the manifold.} and (\ref{grad2}) which should be
supplemented with weights of genus-expanded multi-resolvents under
the first transformation, (\ref{grad1}): $\deg
\rho^{(p|k)}=-3+\frac{5}{2}k+3p$, and under the second
transformation, (\ref{grad2}): $\deg'\rho^{(p|k)}=2-k-2p$.

All the proofs in this subsection are done by induction. Note that
we often omit as trivial checking the induction base. An equivalent
way to obtain multi-resolvents is described in s.\ref{aproofs}.

\paragraph{Proof of (\ref{rho_p_m}): \label{prop_g}}
  Using (\ref{rec_rel_s}) it is easy to show by induction that
$\rho^{(p|m)}(\{z_i\})$ is meromorphic on the $\mathbb{CP}^1$ in
each $z_i$ and can have poles only at $Y=0$. Then, in order to prove
(\ref{rho_p_m}), it is enough to show (due to the symmetricity in
$\{z_i\}$) that for some $z_j$ the order of pole in $Y=0$ is
$(6p+2m-3)$. This also can be done straightforwardly by induction.
Then the degree of the polynomial $Q_{p,m}$ is completely fixed by
symmetry (\ref{grad1}): $\deg Q_{p,m}=(m-1)(m+3p-3)$, while the
power of $M$ in (\ref{rho_p_m}) can be determined by symmetry
(\ref{grad2}). $\square$

In fact, the second symmetry (\ref{grad2}) allows us to put
hereafter $M=1$, while the first symmetry (\ref{grad1}) does not
affect $M$ at all.

\paragraph{Proof of (\ref{rho_0_k_g}): \label{prop_rho_0_k_g}}
 Without any loss of generality one can put $s=0$ (i.e.
$y(z)=z^{1/2}$) using formula (\ref{97}) and identify
$\d_s\equiv-\sum_{i=0}^k\d_{z_i}$. From the definition of $\rho$'s
and using symmetry (\ref{grad1}), one obtains \be
\rho^{(0|k)}(\{z_i\}_{i=1}^k)=\sum_{\sum n_i=k-3}\frac{C_{n_1\ldots
n_k}}{\prod_{i=1}^kz^{3/2+n_i}} \ee \ni In each term of the sum at
the r.h.s. of this expression there exists $i$ such that $n_i=0$.
Let us denote through the subscript $A$ the coefficient in front of
$\frac{1}{z^{3/2}}\ $ in the asymptotic expansion at
$z\rightarrow\infty$. Since $\rho$ is symmetric in $z_i$'s, \be
\rho_A^{(0|k)}(\{z_i\}_{i=1}^{k-1})\equiv\sum_{\sum
n_i=k-3}\frac{C_{n_1\ldots n_{k-1}0}}{\prod_{i=1}^{k-1}z^{3/2+n_i}}
\ee contains the same information as $\rho^{(0|k)}$ itself. Thus,
$\rho$ can be in principle restored from $\rho_A$. However, we do
not need to do this explicitly. Just assume (\ref{rho_0_k_g}) is
correct for $k\le K$ (it is trivially correct for $K=2$). To prove
it for $k=(K+1)$, it is enough to show that \be
\rho^{(0|K+1)}_A(\{z_i\}_{i=1}^K)=\left(-2\sum_{i=1}^{K}\d_{i}\right)^{K-2}\frac{1}{\prod_{i=1}^{K}
z_i^{3/2}} \ee One can easily do it by considering $\sim
\frac{1}{z}$ term of the asymptotic of the recursive relation
(\ref{rec_rel_s}): \be
\rho^{(0|K+1)}_A(\{z_i\}_{i=1}^K)=\left(-2\sum_{i=1}^{K}\d_{i}\right)\rho^{(0|K)}(\{z_i\}_{i=1}^K)\label{asympt_rec_rel}
\ee \ni Then, (\ref{rho_0_k_g}) is correct by induction. $\square$

\paragraph{Proof of (\ref{rho_1_k_g_connected}):
\label{rho_1_k_g}} Similarly to the genus zero case, put $s=0$ and
write \be \rho^{(1|k)}(\{z_i\}_{i=1}^k)=\sum_{\sum
n_i=k}\frac{C_{n_1\ldots n_k}}{\prod_{i=1}^kz_i^{3/2+n_i}} \ee Now
either there exists such $i$ that $n_i=0$ or $\forall i\;n_i=1$.
Therefore, in contrast with the genus zero case, all the information
about $\rho^{(1|k)}$ is contained both in \be
\rho_A^{(1|k)}(\{z_i\}_{i=1}^{k-1})\equiv\sum_{\sum
n_i=k}\frac{C_{n_1\ldots n_{k-1}0}}{\prod_{i=1}^{k-1}z_i^{3/2+n_i}}
\ee \ni {\it and} in $c_k\equiv 3^{-k} C_{1\ldots 1}\ $. First, we
deal with this $c_k$. One can easily construct such $\phi$ that
$\oint_{C}dx\frac{\phi(x)}{x^{n+5/2}}=\frac{\alpha}{3}\delta_{n,0}$.
Now, we are interested only in the terms where each $\d_s$ acts once
on each $\frac{1}{y(z_i)^3}$ in the k-point function in
(\ref{rho_1_k_g_rough}). Then, (\ref{rho_1_k_g_rough}) reads as \be
G[\phi]=\sum_{k=0}^{\infty}\frac{\alpha^k}{k!}\cdot
k!=\frac{1}{1-\alpha} \ee and \be \log G[\phi]=-\log
(1-\alpha)=\sum_{k=1}^{\infty}\frac{\alpha^k}{k}\ee i.e.,
(\ref{rho_1_k_g_connected}) is equivalent to the equality \be
c_k=\frac{(k-1)!}{24} \ee  Now, as usual, we prove it by induction.
Assume that, indeed, $c_k=(k-1)!/24$ for $k\le K$. Computing the
term $\frac{1}{z^2\prod_{i=1}^Kz_i^{3/2}}$ of the asymptotic of
(\ref{rec_rel_s}), one obtains (again only the first term at the
r.h.s. survives) \be
 3^{K+1}c_{K+1}=-2K(3^Kc_K)+5K(3^kc_K)=3^{K+1}Kc_K
\ee where the two terms come from differentiating the denominator
$(z-z_i)$ and the numerator respectively (one then expands
$\frac{1}{z-z_i}\d_{z_i}=\frac{1}{z}\d_{z_i}+\frac{1}{z^2}z_i\d_{z_i}+\ldots$).
Therefore, $c_K=\frac{(K+1)!}{24}\ $.

As the second step, we study $\rho^{(1|k)}_A$. The
$\frac{1}{z}$-asymptotic of the (\ref{rec_rel_s}) gives us \be
\rho^{(1|k+1)}_A(\{z_i\}_{i=1}^k)=\left(-2\sum_{i=1}^{k}\d_{i}\right)\rho^{(1|k)}(\{z_i\}_{i=1}^k).
\label{rho_1_k_A} \ee \ni Introduce the notation
$\tilde{\rho}^{(k)}(\{z_i\}_{i=1}^k)\equiv(2\d_s)^k\frac{1}{\prod_{j=1}^ky^3(z_j)}$
and $\tilde{\rho}^{(k)}_{\tt[conn]}(\{z_i\}_{i=1}^k)$ for its
connected part. Now we again apply the induction and assume that,
for $k\le K$,
$\rho^{(1|k)}(\cdot)=\frac{1}{24}\tilde{\rho}^{(k)}_{\tt[conn]}(\cdot)$.
To prove it for $k=K+1$ it is enough to show that \be
\rho^{(1|K+1)}_A(\cdot)=\frac{1}{24}\tilde{\rho}^{(K+1)}_{\tt[conn],\,A}(\cdot)
\ee \ni Due to (\ref{rho_1_k_A}) and to the induction assumption, it
is equivalent to \be
\tilde{\rho}^{(K+1)}_{\tt[conn],\,A}(\cdot)=2\d_s\tilde{\rho}^{(K)}_{\tt[conn]}(\cdot)
\ee For the complete functions such a relation is obvious from the
definition:
$\tilde{\rho}^{(K+1)}_{A}(\cdot)=2\d_s\tilde{\rho}^{(K)}(\cdot)\ $.
Let $I={1\ldots k}$ and $z_0\equiv z$. The complete $(K+1)$-point
function is expressed through the connected ones as follows
\[
\tilde{\rho}^{(K+1)}(\{z_i\}_{i\in
I\cup\{0\}})=\sum_{\bigsqcup\limits_{j=1}^s\tilde{I}_j=I\cup\{0\}}\prod_{j=1}^s\tilde{\rho}_{\tt[conn]}^{(|\tilde{I}_j|)}(\{z_i\}_{i\in
\tilde{I}_j})=\]
\be=\sum_{\bigsqcup\limits_{j=1}^s{I}_j=I}\sum_{l=1}^s\prod_{\parbox{7mm}{\vspace{-4mm}\center\tiny
$j=1$\\$j\neq
l$}}^s\tilde{\rho}_{\tt[conn]}^{(|{I}_j|)}(\{z_i\}_{i\in
\tilde{I}_j})\tilde{\rho}_{\tt[conn]}^{(|{I}_l|+1)}(\{z_i\}_{i\in
{I}_l\cup \{0\}}) \ee The $\frac{1}{z}$-asymptotic of this equation
is \be\tilde{\rho}^{(K+1)}_A(\{z_i\}_{i\in
I})=\sum_{\bigsqcup\limits_{j=1}^s{I}_j=I}\sum_{l=1}^s\prod_{\parbox{7mm}{\vspace{-4mm}\center\tiny
$j=1$\\$j\neq
l$}}^s\tilde{\rho}_{\tt[conn]}^{(|{I}_j|)}(\{z_i\}_{i\in
\tilde{I}_j})\cdot\tilde{\rho}_{\tt[conn],\,A}^{(|{I}_l|+1)}(\{z_i\}_{i\in
{I}_l}) \ee \ni Now, using the induction assumption, \be
2\d_s\tilde{\rho}^{(K)}(\{z_i\}_{i\in
I})=\tilde{\rho}_{\tt[conn],\,A}(\{z_i\}_{i\in
I})+\sum_{\bigsqcup\limits_{j=1}^{\mathbf{s>1}}{I}_j=I}\sum_{l=1}^s\prod_{\parbox{7mm}{\vspace{-4mm}\center\tiny
$j=1$\\$j\neq
l$}}^s\tilde{\rho}_{\tt[conn]}^{(|{I}_j|)}(\{z_i\}_{i\in
\tilde{I}_j})\cdot2\d_s\tilde{\rho}_{\tt[conn]}^{(|{I}_l|)}(\{z_i\}_{i\in
{I}_l}) \ee \be \Leftrightarrow\hspace{1cm}
2\d_s\sum_{\bigsqcup\limits_{j=1}^{s}{I}_j=I}\prod_{j=1}^s\tilde{\rho}_{\tt[conn]}^{(|{I}_j|)}(\{z_i\}_{i\in
\tilde{I}_j})=\tilde{\rho}_{\tt[conn],\,A}(\{z_i\}_{i\in
I})+2\d_s\sum_{\bigsqcup\limits_{j=1}^{\mathbf{s>1}}{I}_j=I}\prod_{j=1}^s\tilde{\rho}_{\tt[conn]}^{(|{I}_j|)}(\{z_i\}_{i\in
\tilde{I}_j}) \ee \be
\Longrightarrow\hspace{5mm}\tilde{\rho}^{(K+1)}_{\tt[conn],\,A}(\{z_i\}_{i\in
I})=2\d_s\tilde{\rho}^{(K)}_{\tt[conn]}(\{z_i\}_{i\in
I})\hspace{5mm}\Longrightarrow\hspace{5mm}
\rho^{(1|K+1)}_A(\{z_i\}_{i\in
I})=\frac{1}{24}\tilde{\rho}^{(K+1)}_{\tt[conn],\,A}(\{z_i\}_{i\in
I}) \ee \ni $\square$

\subsection{Matrix integral representation}

This partition function $Z_K$ can be presented as the Hermitian
matrix integral depending on the external matrix $A$, \be
\label{Kint}Z_K= {\int DX\ \exp\left(-{g^2\over
3}\hbox{Tr}X^3-{g\over\sqrt{3}}\hbox{Tr}AX^2\right)\over \int DX\
\exp\left(-{g\over\sqrt{3}}\hbox{Tr}AX^2\right)}\ee where the
integral is understood as a perturbative power series in
$\tau_{2k+1}\equiv \displaystyle{g}{3^{2k+1}\over 2k+1}\hbox{Tr}
A^{-2k-1}$. Note that this integral does not depend on the size of
matrices $X$ and $A$ provided it is being considered as a function
of $t_k$ \cite{GKM}. By the shift of the integration variable, it
can be also reduced to the form \be Z_K= \exp\left(-{2\over
3g}\hbox{Tr}\Lambda^{3\over 2}\right){\int DX\ \exp\left(-{g^2\over
3}\hbox{Tr}X^3+\hbox{Tr}\Lambda X\right)\over \int DX\
\exp\left(-{g\over\sqrt{3}}\hbox{Tr}AX^2\right)} \ee where
$3\Lambda=A^2$, i.e. $\tau_{2k+1}\equiv \displaystyle{g{1\over
2k+1}\Tr \Lambda^{-k-{1\over 2}}}$\ .\footnote{In order to introduce
an arbitrary shifted first time, $t_{2k+1}=\tau_{2k+1}-{M\over
3}\delta_{k,1}$, where $M$ is a parameter, one should consider
instead of (\ref{Kint}) the integral \be Z_K= {\int DX\
\exp\left(-{16g^2\over
3M^2}\hbox{Tr}X^3-{2\sqrt{2}g\over\sqrt{3M}}\hbox{Tr}AX^2\right)\over
\int DX\ \exp\left(-{2\sqrt{2}g\over\sqrt{3M}}\hbox{Tr}AX^2\right)}=
\exp\left(-{M\over 6g}\hbox{Tr}\Lambda^{3\over 2}\right){\int DX\
\exp\left(-{16g^2\over 3M^2}\hbox{Tr}X^3+\hbox{Tr}\Lambda
X\right)\over \int DX\
\exp\left(-{2\sqrt{2}g\over\sqrt{3M}}\hbox{Tr}AX^2\right)}\nn\ee
which is a function of the same
$\tau_{2k+1}=g\displaystyle{{3^{k+{1\over 2}}\over k+{1\over
2}}\hbox{Tr} A^{-2k-1}}=\displaystyle{{g\over k+{1\over 2}}\hbox{Tr}
\Lambda^{-k-{1\over 2}}}$.}

\subsection{Okounkov's representation of the Laplace transformed resolvents}
Further on in this section we put $M=2$, $g=1$ and $s=0$.

\subsubsection{Laplace transform of the resolvents}
Let us introduce the Laplace-transformed resolvents $\eta$:
\begin{equation}
 \rho^{|k)}(z_1,\ldots,z_k)=2^k\,\int\limits_0^\infty\prod_{i=1}^kdx_i e^{-\sum\limits_i x_iz_i}\eta^{|k)}(x_1,\ldots,x_k) \label{rho_eta}
\end{equation}
\begin{equation}
 \eta^{|k)}(x_1,\ldots,x_k)=\frac{1}{(4\pi\,i)^k}\oint\limits_C\prod_{i=1}^kdz_i e^{\sum\limits_i x_iz_i}\rho^{|k)}(z_1,\ldots,z_k) \label{eta_rho}
\end{equation}
where the contour $C$ encircles 0, beginning and ending at the
negative infinity with respect to the branch cut along the negative
real ray.

The manifest expressions for a few first $\eta$ are
\begin{equation}
 \eta^{(0|3)}(x_1,x_2,x_3)=\frac{1}{2\,\pi^{3/2}}\,x_1^{1/2}x_2^{1/2}x_3^{1/2}
\end{equation}
\ba
 \eta^{(0|4)}(x_1,x_2,x_3,x_4)=\frac{1}{2\,\pi^2}\,(x_1^{1/2}x_2^{1/2}x_3^{1/2}x_4^{3/2}+x_1^{1/2}x_2^{1/2}x_3^{3/2}x_4^{1/2}\\
+x_1^{1/2}x_2^{3/2}x_3^{1/2}x_4^{1/2}+x_1^{3/2}x_2^{1/2}x_3^{1/2}x_4^{1/2})
\ea
\begin{equation}
 \eta^{(1|1)}(x_1)=\frac{1}{24\,\pi^{1/2}}\,x_1^{3/2}
\end{equation}
\begin{equation}
 \eta^{(1|2)}(x_1,x_2)=\frac{1}{24\,\pi}\,(x_1^{1/2}x_2^{5/2}+x_1^{5/2}x_2^{1/2}+x_1^{3/2}x_2^{3/2})
\end{equation}
\ba
 \eta^{(1|3)}(x_1,x_2,x_3)=\frac{1}{24\,\pi^{3/2}}\,(x_1^{1/2}x_2^{1/2}x_3^{7/2}+x_1^{1/2}x_2^{7/2}x_3^{1/2}+x_1^{7/2}x_2^{1/2}x_3^{1/2}+2x_1^{1/2}x_2^{3/2}x_3^{5/2}\\+2x_1^{1/2}x_2^{5/2}x_3^{3/2}+2x_1^{3/2}x_2^{1/2}x_3^{5/2}+2x_1^{5/2}x_2^{1/2}x_3^{3/2}+2x_1^{3/2}x_2^{5/2}x_3^{1/2}\\+2x_1^{5/2}x_2^{3/2}x_3^{1/2}+2x_1^{3/2}x_2^{3/2}x_3^{3/2})
\ea
\begin{equation}
 \eta^{(2|1)}(x_1)=\frac{1}{2^{10}\,9\,\pi^{1/2}}\,x_1^{9/2}
\end{equation}
\ba
 \eta^{(2|2)}(x_1,x_2)=\frac{1}{2^6\,45\,\pi}(5x_1^{1/2}x_2^{11/2}+5x_1^{11/2}x_2^{1/2}+15x_1^{3/2}x_2^{9/2}\\+15x_1^{9/2}x_2^{3/2}+29x_1^{5/2}x_2^{7/2}+29x_1^{7/2}x_2^{5/2})
\ea One can easily turn on nonzero $s$ using
$\eta^{(p|m)}(x_1,\ldots,x_m|s)=e^{s\sum\limits_i
x_i}\eta^{(p|m)}(x_1,\ldots,x_m|0)$ (except for $p=0$,$m=1$ and
$p=0$,$m=2$ cases).

The general formula for the genus zero $\eta$-resolvents can be
easily obtained from (\ref{rho_0_k_g}): \ba
 \eta^{(0|k)}(x_1,\ldots,x_k)=\frac{1}{(4\pi\,i)^k}\oint\limits_C\prod_{i=1}^kdz_i e^{\sum\limits_i x_iz_i}\frac{1}{2}\,\left(-\sum_{i=1}^k\d_{z_i}\right)^{k-3}\,\frac{1}{z_1^{3/2}\cdots z_k^{3/2}}=\\
=\frac{\left(\sum_{i=1}^k{x_i}\right)^{k-3}}{2\,(4\pi\,i)^k}\oint\limits_C\prod_{i=1}^kdz_i
e^{\sum\limits_i x_iz_i}\,\frac{1}{z_1^{3/2}\cdots
z_k^{3/2}}=\frac{1}{2\,\pi^{k/2}}\left(\sum_{i=1}^k{x_i}\right)^{k-3}\prod_{i=1}^kx_i^{1/2}
\ea

\subsubsection{One point function}\label{sect_eta_one_point}
One can easily verify that the one-point $\eta$-resolvent is
as follows
\begin{equation}
 \eta^{|1)}(x)=\frac{1}{2\,\sqrt{\pi}}\,\frac{e^\frac{x^3}{12}}{x^{3/2}}
\end{equation}
The r.h.s. of (\ref{rho_eta}) is then equal to \ba
2\int\limits_0^\infty dx\,\eta(x)e^{-zx}=\frac{1}{\sqrt{\pi}}\sum_{p=0}^\infty\frac{1}{p!\,12^p}\int\limits_0^\infty dx x^{3p-3/2}e^{-xz}=\\
=\frac{1}{\sqrt{\pi}}\sum_{p=0}^\infty
\frac{1}{p!\,12^p}\,\frac{\Gamma(3p-1/2)}{z^{3p-1/2}}=
\sum_{p=0}^\infty
\frac{(6p-3)!!}{p!\,12^p\,2^{3p-1}}\,\frac{1}{z^{3p-1/2}} \ea
Exactly the same expression is obtained from (\ref{rho_p_1_expansion})
\begin{equation}
 \rho^{|1)}=\sum_{p=0}^\infty \rho^{(p|1)}(x)=\sum_{p=0}^\infty \frac{(6p-3)!!}{p!\,24^p\,2^{2p-1}}\,\frac{1}{z^{3p-1/2}}
\end{equation}

\subsubsection{Okounkov's result}
In \cite{ok} Okounkov obtained a representation for the resolvents
in terms of finite-dimensional integrals which can be
thought of as a discrete version of a special functional integral.
He introduced the functions \be\label{forE} \cE(x_1,\dots,x_n)=
\frac1{2^n\pi^{n/2}} \frac{\exp\left(\frac1{12} \sum x_i^3\right)
}{\prod \sqrt{x_i}}\int_{s_i\ge 0} ds \, \exp\left(-\sum_{i=1}^n
\frac{(s_i-s_{i+1})^2}{4x_i} -\sum_{i=1}^n \frac{s_i+s_{i+1}}2\,x_i
\right) \ee and their symmetrized versions
\begin{equation}
 \cEt(x) =
\sum_{\sigma\in S(n)/(12\dots n)}
\cE(x_{\sigma(1)},\dots,x_{\sigma(s)}) \,,
\end{equation}
where the summation is over coset representatives modulo the cyclic
group generated by the permutation $(12\dots n)$.

Then the resolvent is a generating function for intersection numbers on moduli spaces
of curves with $n$ fixed points (genus is arbitrary) and is equal (up to
some renormalisation and rescaling) to the sum\footnote{In
\cite{ok} Okounkov used notation $\cG$ instead of $\eta$.}
\begin{equation}
 \eta^{|n)}(x_1,\dots,x_n) = \sum_{\alpha\in\Pi_n} (-1)^{\ell(\alpha)+1} \,  \cEt(x_\alpha)\,,\label{eta_okounkov}
\end{equation}
where $\Pi_n$ is the set of all partitions $\alpha$ of the set
$\{1,\dots,n\}$ into disjoint union of subsets. For any partition
$\alpha\in\Pi_n$ with $\ell=\ell(\alpha)$ blocks, $x_\alpha$ is the
vector of size $\ell$ formed by sums of $x_i$ over the blocks of
$\alpha$.

For $n=1$ formula (\ref{eta_okounkov}) is very simple:
$\eta^{|1)}(x)=\cEt(x)=\cE(x)=\frac{1}{2\,\sqrt{\pi}}\,\frac{e^\frac{x^3}{12}}{x^{3/2}}$
and it is exactly what we obtained in section
\ref{sect_eta_one_point}.

Note that Okounkov used the time variables that differ from those typically used
in KdV by the factor of $(2k+1)!!$ for the $(2k+1)$th
time variable\footnote{We do not care here about some $k$-independent
factor, because it can be easily eliminated by rescaling.}.
In order to reproduce these factors in the definition of the resolvent,
one has to use here the Laplace-transformed resolvent $\eta (x)$
instead of $\rho (z)$, since under the Laplace transform
\be
x^{k+1/2}\longrightarrow {2^{k+1}\over\sqrt{\pi}(2k+1)!!}{1\over z^{k+3/2}}
\ee

\subsubsection{$L_{-1}$ Virasoro constraint and genus-zero resolvents}

In this section we show explicitly that Okounkov's functions
(i.e. the r.h.s. of (\ref{eta_okounkov})) satisfy the lowest $L_{-1}$ Virasoro
constraint. Note that, when proving (\ref{rho_0_k_g}) in sect.\ref{prop_rho_0_k_g},
we used only the leading asymptotics of
the recursive relations in $z$, which is equivalent to the $L_{-1}$-constraint.
In fact, the $L_{-1}$-constraint is sufficient to determine
the genus zero resolvents in the case when only $T_3\neq 0$. Hence by
demonstrating the function satisfies the $L_{-1}$-constraint one
is automatically guaranteed the genus zero result is correct (therefore,
it is possible just to apply the same arguments as in proof of (\ref{rho_0_k_g})).

The $L_{-1}$-constraint imposed on the partition function is equivalent to the following constraint
on the asymptotics of the $\rho$-resolvents (compare with (\ref{asympt_rec_rel})):
\begin{equation}\label{rhores}
 \rho^{|n)}(z_1,\ldots,z_{n})=-\frac{2}{z_n^{3/2}}\left(\sum_{i=1}^{n-1}\d_i\right)\rho^{|n-1)}(z_1,\ldots,z_{n-1})+o(z_n^{-3/2}),\;\;z_n\rightarrow\infty
\end{equation}
Under the Laplace transform, (\ref{eta_rho}) it leads to
 \begin{equation}
  \eta^{|n)}(x_1,\cdots,x_n)=\sqrt{\frac{x_n}{\pi}}\,(x_1+\cdots+x_{n-1})\,\eta^{|n-1)}(x_1,..,x_{n-1})+o(\sqrt{x_n}),\;\;\;x_n\rightarrow 0 \label{eta_asympt}
 \end{equation}
i.e. the $\eta$-resolvents given by formula (\ref{eta_okounkov})
satisfy the $L_{-1}$ constraint.

In order to prove this formula, one needs to use the following identity
\ba\label{lemma_asympt_ok}
 \int\limits_0^\infty da\int\limits_0^\infty db\, f(a,b)\,\frac{e^{-\frac{(a-b)^2}{2\epsilon}}}{\sqrt{2\pi\epsilon}}=
\\
 =\int\limits_0^\infty dc\,f(c,c)-\sqrt{\frac{\epsilon}{2\pi}}\,f(0,0)+O(\epsilon)\equiv\int\limits_0^\infty dc\left\{\,f(c,c)+\sqrt{\frac{\epsilon}{2\pi}}\,\d_c f(c,c)\right\}+O(\epsilon)
\ea
It follows from the decomposition
\begin{equation}
 f(a,b)=\left[f(a,b)-\Theta(\Lambda-(a+b))f(0,0)\right]+f(0,0)\,\Theta(\Lambda-(a+b))
\end{equation}
where $\Theta$ is the Heaviside step function. The difference in the brackets can be
proved not to contain
the $O(\sqrt{\epsilon})$ term, while the remaining integral can be
computed exactly.

Now, one can apply to the Laplace transform,
(\ref{eta_rho}) of the $\rho$-resolvents (\ref{rhores})
formula (\ref{lemma_asympt_ok}) with $\epsilon=2x_k$
 \begin{multline}
\cE(x_1,\dots,x_n)= \frac{1}{2^n\pi^{n/2}}
\frac{\exp\left(\frac1{12} \sum x_i^3\right)
}{\prod \sqrt{x_i}}\times \\
\int\limits_0^\infty\,\prod_{i=1}^n ds_i\,
e^{\left\{\cdots-\frac{(s_{k-1}-s_k)^2}{4x_{k-1}}-\frac{(s_{k}-s_{k+1})^2}{4x_{k}}-
\frac{(s_{k+1}-s_{k+2})^2}{4x_{k+1}}-\cdots-s_{k}\frac{x_k+x_{k-1}}{2}-s_{k+1}\frac{x_{k+1}+x_{k}}{2}\cdots
\right\}}\stackrel{=}{\hbox{\tiny{$x_k\rightarrow 0$}}}
\end{multline}

\begin{equation}
 =\frac{1}{2^{n-1}\pi^{(n-1)/2}}
\frac{\exp\left(\frac1{12} \sum\limits_{i\neq k} x_i^3\right)
}{\prod\limits_{i\neq k}
\sqrt{x_i}}\int\limits_0^\infty\,\prod_{i\neq k,k+1} ds_i\,ds\,
A\,e^{\left\{\cdots-\frac{(s_{k-1}-s)^2}{4x_{k-1}}-\frac{(s-s_{k+2})^2}{4x_{k+1}}-\cdots-s\frac{x_{k+1}+x_{k-1}}{2}\cdots
\right\}}
\end{equation}
where
\begin{equation}
 A=1+\sqrt{\frac{x_k}{\pi}}\,\d_{s}+o(\sqrt{x_k})=1-\sqrt{\frac{x_k}{\pi}}\left[\frac{s-s_{k+2}}{2x_{k+1}}-\frac{s_{k-1}-s}{2x_{k-1}}+\frac{x_{k-1}+x_{k+1}}{2}\right]+o(\sqrt{x_k})
\end{equation}
In $\cEt$ some terms cancels
\begin{equation}
\cEt(x_1,\cdots,x_n)=\left[(n-1)-\sqrt{\frac{x_n}{\pi}}(x_1+\cdots+x_{n-1})\right]\cEt(x_1,\cdots,x_{n-1})+o(\sqrt{x_n})
\end{equation}
Note that if $\alpha$ is a partition (see \cite{ok}), then there are two possibilities: 1) it
does not contain the block $\{x_n\}$, then
\begin{equation}
 \cEt(x_\alpha)=\cEt(x_\alpha|_{x_n=0})+o(\sqrt{x_n})
\end{equation}
2) it contains this block, then
\begin{equation}
 \cEt(x_\alpha)=\left[(\ell(\alpha)-1)-\sqrt{\frac{x_n}{\pi}}(x_1+\cdots+x_{n-1})\right]\cEt(x_{\alpha\backslash\{x_n\}})+o(\sqrt{x_n})
\end{equation}
In $\eta$ further cancelations take place so that finally one arrives at formula
(\ref{lemma_asympt_ok}).

As it was already discussed in the beginning of this subsubsection, one can easily derive
using (\ref{eta_asympt}) that
\begin{equation}
 \eta^{(0|k)}(x_1,\ldots,x_k)=\frac{1}{2\,\pi^{k/2}}\left(\sum_{i=1}^k{x_i}\right)^{k-3}\prod_{i=1}^kx_i^{1/2}
\end{equation}
Verification of the $L_0$-constraint needs much enhanced
version of (\ref{lemma_asympt_ok}) and much more involved computations.

\section{Simplest DV type solution to Kontsevich model: KdV hierarchy} \label{KdV_section}
\setcounter{equation}{0}

As we already discussed in sect. 2, there are many solutions to the generic
Kontsevich model parameterized by an arbitrary function $F[T]$ that
satisfies two constraints (\ref{init_constr}), in variance with
the Gaussian case of the previous section, when the solution is
unique. However, among all these many solutions there is a special
family of the so-called Dijkgraaf-Vafa solutions. They are
associated with a Riemann surface which genus is generically equal
to $N-1$, the number of non-zero times $T_k$ being equal to $N+1$.
These solutions are non-generic,
and, being extended to depend on higher times
$T_k$, $k>2N+1$, are required to be associated with the same Riemann surface and to
have a smooth limit upon bringing these excessive
times to zero in order \cite{mir}. Moreover, one typically
considers infinitely many excessive times (still keeping the genus $N-1$ of the curve
fixed). Then, $F^{(0)}[T]$ is logarithm of the $\tau$-function of a
Whitham hierarchy w.r.t. these infinitely many times $T_k$
\cite{ChM,CMMV}, while the complete matrix model partition function
as a function of $T_k$ corresponds to a dispersionful integrable
hierarchy.

In this section we consider the simplest DV solution, that is, the
solution associated with a sphere. The partition function $Z_K(t)$
(as well as $Z_K[T]\equiv Z_K(t)|_{\tau=0}$) of this system is nothing but a
$\tau$-function of the KdV hierarchy, while its planar limit,
$F^{(0)}[T]$ is logarithm of the $\tau$-function of the
dispersionless KdV hierarchy (i.e. Whitham hierarchy in the case of
spherical Riemann surface). This is exactly the solution that was
previously considered as relevant to $2d$ gravity
\cite{Douglas,FKN1,DVV,GKM,versus}.

Throughout this section we denote
\be
U=\frac{\p^2{F}}{\p T_1^2}
\ee
and its dispersionless counterpart
\begin{equation}\label{dc}
 u=\frac{\p^2{F^{(0)}}}{\p T_1^2}
\end{equation}
 The Lax operator is
\be {\bf L}=g^2\d^2+2U
\ee
The evolution is given by the flows
\begin{equation}\label{dc2}
 \frac{\d {\bf L}}{\d T_{i}}=g[{\bf L}^{\frac{i}{2}}_+,{\bf L}]
\end{equation}
Here $[\cdot]_+$ denotes the differential part of the pseudo-differential
operator (i.e. the "non-negative part" in the formal operator $\d$).
The appropriate solution of the Virasoro constraints is fixed by
the string equation
\begin{equation}
[{\bf L},{\bf M}]=2g  \label{Douglas}
\end{equation}
where
\be
{\bf M}=\sum_{k=1}^N(2k+1)T_{2k+1}\,{\bf L}^{k-1/2}_+
\ee

\subsection{An example: $\mathbf{N=2}$ case}\label{subsect_KdV_N_2}

Let us assume that $Z[T]=e^{F[T]/g^2}$, which we ''found``
(up to some arbitrary function) in subsection \ref{sect_RVC} solving the
first Virasoro constraints, is, in addition, a $\tau$-function of the KdV (or
the 2-reduced KP) hierarchy. The 3rd and the 5th equations of the hierarchy are
\ba \frac{1}{3}\frac{\d U}{\d T_3}=\frac{\d}{\d
T_1}\left(\frac{U^2}{2}+g^2\frac{U''}{12}\right)\\
\frac{1}{15}\frac{\d U}{\d T_5}=\frac{\d}{\d
T_1}\left(\frac{U^3}{6}+g^2\frac{UU''}{12}+g^2\frac{U'^2}{24}+g^4\frac{U''''}{240}\right)
\ea

\ni Picking up $F$ from (\ref{red_sol_N2}) and inserting it into these equations leads to
the Painlev\'e-I equation on $G(\eta_2)\equiv \tilde{\tilde F}''(\eta_2)$:
\be  g^2G''=-6\cdot G^2+\frac{3^55^4}{4}\eta_2 \label{Painvele}\ee

\ni Solving it perturbatively w.r.t. $g^2$, one obtains
series (\ref{elliptic_g_exp}) with the coefficients being recursively
determined\footnote{Note that, strictly speaking, there are two solutions to (\ref{Painvele})
corresponding to the two branches of $\eta_2^{5/2}$ in (\ref{elliptic_g_exp})
(or, equivalently, to the two branches of $\sqrt{T_3^2-\frac{10}{3}T_1T_5}$ in the expressions
below). But only one of these branches has a smooth limit $T_5\rightarrow 0$,
i.e. permits the transition $N=2\rightarrow N=1$ considered in subsection \ref{ell_rat}.
Therefore, only one of them is associated with the DV solution.}.
The first two coefficients coincide with those in
(\ref{asympt_eta_0}). From that equation it follows that
$\tilde{\tilde F}$ may also have an extra term $C'\eta_2$. However, one can
check that actually $C'=0$ using e.g. the equation
\be \res {\bf L}^{5/2}=\frac{\d^2{F}}{\d T_1\d T_5} \ee
where the residue is defined by
\be
\res \sum_{k=-\infty}^\infty a_k(g\p)^k=a_{-1}
\ee

Thus, we have fixed all $F^{(p)}[T]$ completely, there is
no more ambiguity and the curve and the densities are fixed. As we
mentioned, ${F}^{(0)}[T]$ is such that the elliptic curve actually
degenerates:

\be
y^2=\left({5}T_5\right)^2\left(x+\frac{1}{5}\frac{\sqrt{T_3^2-\frac{10}{3}T_1T_5}+2T_3}{T_5}\right)^2\left(x-\frac{2}{5}\frac{\sqrt{T_3^2-\frac{10}{3}T_1T_5}-T_3}{T_5}\right)
\ee

\ni i.e. the torus is pinched at the point
$y=0,\;x=-\frac{1}{5}\frac{\sqrt{T_3^2-\frac{10}{3}T_1T_5}+2T_3}{T_5}$.
Thus the curve is equivalent to the rational one
$Y^2=x-\frac{2}{5}\frac{\sqrt{T_3^2-\frac{10}{3}T_1T_5}-T_3}{T_5}$ .

A few first (multi)-densities are (here we consider the critical
point\footnote{By critical point we mean the case when
$T_{2N+1}=\mathrm{const}$ and other $T_i=0,\;i>1$. $T_1$ is as usual a variable.} case
$T_3=0,\;T_5=\frac{-2}{15}$ to make formulae more compact):

\be
y(z)=-\frac{2}{3}(\sqrt{T_1}+z)Y(z),\;\;\;\;Y^2(z)=z-2\sqrt{T_1}\ee

\be
\rho^{(0|2)}(z_1,z_2)=-\frac{4\sqrt{T_1}-z_1-z_2+\frac{z_1+z_2}{\sqrt{z_1z_2}}Y(z_1)Y(z_2)}{Y(z_1)Y(z_2)(z_1-z_2)^2}
=\left.\rho^{(0|2)}_{\hbox{\tiny{(N=1)}}}(z_1,z_2)\right|_{s=2\sqrt{T_1}}\ee

\be
\rho^{(0|3)}(z_1,z_2,z_3)=\frac{1}{2\sqrt{T_1}}\frac{1}{Y^3(z_1)Y^3(z_2)Y^3(z_3)}\ee

\small
\begin{equation*}
 \rho^{(0|4)}(z_1,z_2,z_3,z_4)=-\frac{1}{4\,T_1^{3/2}{Y^5(z_1)Y^5(z_2)Y^5(z_3)Y^5(z_4)}}(z_1z_2z_3z_4-5\,T_1^{1/2}(z_1z_2z_3+z_1z_2z_4+z_1z_3z_4+z_2z_3z_4)+
\end{equation*}
\begin{equation}
 +16\,T_1(z_1z_2+z_1z_3+z_1z_4+z_2z_3+z_2z_4+z_3z_4)-44\,T_1^{3/2}(z_1+z_2+z_3+z_4)+112\,T_1^2)
\end{equation}
\normalsize

\be \rho^{(1|1)}(z)=-\frac{1}{48\cdot
T_1}\frac{z-5\sqrt{T_1}}{Y^5(z)}\ee

\be \rho^{(2|1)}(z)=-\frac{7}{9\cdot 2^{10}\cdot
T_1^{7/2}}\frac{613T_1^2-503T_1^{3/2}z+204T_1z^2-44T_1^{1/2}z^3+4z^4}{Y^{11}(z)}
\ee

\be
\rho^{(1|2)}(z_1,z_2)=-\frac{1}{3\cdot
2^7\cdot T_1^4\cdot Y^7(z_1)Y^7(z_2)}\cdot \\
\cdot\left(2z_1^2z_2^2-14\sqrt{T_1}(z_1^2z_2+z_2^2z_1)+35T_1(z_1^2+z_2^2)+89T_1z_1z_2-182T_1^{3/2}(z_1+z_2)+284T_1^2\right) \ee

Thus, we checked that if $Z=e^{\mathcal{F}/g^2}$
satisfies the Virasoro constraints (for $N=2$), then the following statements are equivalent:

$\bullet$ $Z$ is a $\tau$-function of the KdV-hierarchy

$\bullet$ The elliptic curve $y(z)$ degenerates into the rational
one and the multi-densities $\rho^{(p|m)}$ have no poles at the two
marked points on the sphere which come from the double point
singularity on the torus.

Equivalently, one can say that the poles are only at $Y=0$ points.
Thus, the condition imposed on the curve determines $F^{(0)}[T]$ and
then the condition of canceling the singularities determines
$F^{(p)}[T],\;p>0$ (for general $F^{(p)}[T]$ the poles at
$x=-\frac{1}{5}\frac{\sqrt{T_3^2-\frac{10}{3}T_1T_5}+2T_3}{T_5}$ do
exist). We will specify this statement for general case below.

\subsection{Quasiclassical limit of the KdV hierarchy as Whitham hierarchy}\label{sect_eq_to_w}

In this subsection we review some features of the quasiclassical (or
dispersionless) limit of the KdV hierarchy and its representation in
terms of the generalized Whitham hierarchy
\cite{Krichever_Whitham,RG}. We work only quasiclassically and
suppress the corresponding $(0)$ superscript. In the quasiclassical
limit of the KdV hierarchy, the momentum is just a commuting
variable: $g\d \rightsquigarrow P$, because one can neglect all the
commutators $[P,\cdot]\sim g$. Then, the Lax operator is just a
function \be L=P^2+2u\ee

\ni which satisfy an additional constraint -- the string equation:
\be \{L,M\}=2 \label{str} \ee

\ni where $\{\cdot,\cdot\}$ is the Poisson brackets ($\{P,T_1\}=1$)
and \be
M=\sum_{k=1}^N(2k+1)T_{2k+1}\,L^{k-1/2}_+=\sum_{j=0}^{N-1}M_j P^{2j+1}
\label{Q_oper}\ee

By $[\cdot]_+$ we denote the positive part of expansion at the
vicinity of $P=\infty$. $T_k$ and $\frac{\d F}{\d T_k}$ can be
represented as follows (see proof in Appendix \ref{app_disp_KdV}): \be
 T_{k}=-\frac{1}{k}\res_{P=\infty}\left\{P\,M L^{-k/2}dP\right\} \\
 \frac{\d F}{\d T_{k}}=-\res_{P=\infty}\left\{P\,M L^{k/2}dP\right\}
\label{w_res} \ee
This defines our system as generalized Whitham hierarchy for the sphere
parameterized by $P$.

For generic $N$, the curve (\ref{curve}) also degenerates to the rational
one, when one imposes on $Z$ an additional constraint to be the
$\tau$-function of the KdV hierarchy. This is actually an obvious
consequence of the equivalence of the curve (\ref{curve}) and the one
appearing in the Whitham hierarchy, since the dispersionless limit of KdV
(which describes $F^{(0)}[T]$ in (\ref{curve})) corresponds to
the Whitham hierarchy on the sphere. Hence,
\noindent
{\bf imposing on the solution of KdV hierarchy additional Virasoro constraints quasiclassically is equivalent
to total degeneration of the corresponding curve $\mathcal{C}_{2,2N-1}$ given by (\ref{curve}).
This degenerated curve has the global parametrization $z=L(P),\;y=M(P)$.}

Indeed, let us denote
\be x=P^2+2u=L;~~~\tilde{y}=\sum_{k=0}^{N-1}M_{k}P^{2k+1}\label{curve_w_x}\ee

\ni Then
\be
\tilde{y}^2(x)=(x-2u)\left(\sum_{k=0}^{N-1}M_{k}(x-2u)^k\right)^2
\label{curve_whitham}\ee

\ni The curve $C_{2,2N-1}$ defined earlier reads explicitly as

\be
y^2=\frac{1}{z}\left\{\left(\sum_{k=0}^Nz^k(2k+1)T_{2k+1}\right)^2-T_1^2\right\}+2\sum_{m=0}^{N-2}\sum_{k=m+2}^Nz^{k-m-2}(2k+1)T_{2k+1}\frac{\d
F}{\d t_{2m+1}} \label{curve_expl}\ee

\ni To show that $y^2(x)=\tilde{y}^2(x)$, one can use formulae (\ref{w_res})

\be T_k=-\frac{1}{2k}\res_{P=\infty}\left[x^{-k/2}dS\right]
\label{res_y_t} \ee

\be \frac{\d F}{\d
T_k}=-\frac{1}{2}\res_{P=\infty}\left[x^{k/2}dS\right]
\label{res_y_Ft}\ee
where $d S=\tilde{y}d x$. Then, expression (\ref{curve_expl}) can be
combined into the sum

\be
y^2(z)=\frac{1}{4}\sum_{a+b>0}z^{a+b-1}\res_{P=\infty}\left[x^{-a-1/2}dS\right]\res_{P=\infty}\left[x^{-b-1/2}dS\right]
=[y^2_R(z)]_+\ee
\ni where
\be
y_R(z)=-\frac{1}{2}\sum_az^{a-1/2}\res_{P=\infty}\left[x^{-a-1/2}dS\right]=
-\sum_az^{a-1/2}\res_{x=\infty}\left[x^{1/2}\tilde{y}(x)\frac{dx}{x^{a+1}}\right]=\tilde{y}(z)\ee
Therefore
\be y^2(z)=[\tilde{y}^2(z)]_+=\tilde{y}^2(z)\ee

The inverse is also true. The condition of curve degeneration
into sphere gives us $N-1$ independent linear PDEs on
$F$ (pinching of all $N-1$ handles, or coinciding
of the corresponding $N-1$ pairs of roots) which fix completely the ambiguity, the function
$\wt^{(0)}$ of $N-1$ variables.

When all $T_k$'s are fixed, one can completely determine all $M_k$'s and
$u$. That is, $u$ is defined by the equation
\be \sum_{k=0}^N\frac{(2k+1)!!}{k!}\,T_{2k+1}\,u^k=0
\label{string_eq}\ee
which has only $N$ solutions.
Thus, in generic case one obtains\footnote{$Y(x)$ of subsection
\ref{subsect_KdV_N_2} is just $P(x)$.}

\be y(x)=P(x)y_{r}(x)
\;\;\left(P(x)=\sqrt{x-2u},\;\;\;y_r(x)=\sum_{s=0}^{N-1}M_s(x-2u)^s\right)
\label{reduced_curve}\ee

\subsection{An alternative way to obtain multi-resolvents\label{aproofs}}

When $T_1,\ldots,T_{2N+1}$ times are turned on, there are $N$ solutions to the
equations of the KdV hierarchy that satisfy the Virasoro constraints.
The choice of solution corresponds to the choice of a root in equation (\ref{string_eq}).
$N-1$ of these solutions have a smooth limit when $T_{2N+1}\rightarrow 0$,
and one of them (corresponding to the root of (\ref{string_eq}) which goes to infinity) diverges.
However, only one solution survives when all the times
$T_k\rightarrow 0,\;k>3$ reducing in this limit
to the Gaussian model. Putting this differently, one may associate $T_k$'s with $\tau_k$'s of the
Gaussian model in this case.
Partition functions for different choices of $T_k$'s are actually given
by the same function in the sense that
\begin{equation}
 Z(T',\tau')=Z(T,\tau)=Z_K(t)\;\;\;if\;\;\;T'_{2k+1}+\tau'_{2k+1}=T_{2k+1}+\tau_{2k+1}=t_{2k+1}\;.
\end{equation}
However, different solutions, i.e. different functions $F[T_1,\ldots,T_{2N+1}]$,
can be obtained from this one by analytic continuation with respect
to variables\footnote{We use it implicitly in what follows.
E.g., formulae (\ref{mvars-1})-(\ref{mvars-3}) are known for the
Gaussian branch. We derive from them explicit formulas for
$\rho^{(p|k)}$'s, e.g., (\ref{rho_0_k_g-gen}). They are rational
functions in $u$ and $M_k$'s (which in turn are polynomials in $u$
and linear functions in $T_k$'s). Once one have obtained these
formulae for the specific solution to the KdV equations, one can make
analytic continuation in $T_1,\ldots,T_{2N+1}$ to prove that they
hold on over the branches. } $T_1,\ldots,T_{2N+1}$.

In this subsection we explain how to derive formulae analogous to
(\ref{rho_0_k_g}) and (\ref{rho_1_k_g_connected}) for
multi-resolvents for the general DV solution considered in this section.
This derivation is closer to the original way of getting
(\ref{rho_0_k_g}) in \cite{AJKM} (see also \cite{AHW}). The crucial point for the
derivation is to use specific {\it moment} variables
\cite{Itzykson}. The partition function in these variables can be
easily obtained within the realization of the Kontsevich partition
function as the highest weight of the Virasoro algebra given on the
spectral curve (\ref{koncurve}), \cite{AMM.IM}. Within this
approach, one makes a change of the local parameter on the spectral
curve\footnote{Note that the spectral curve considered in
\cite{AMM.IM} is double covering of curve (\ref{koncurve})
considered here. Hence, the local parameter $\xi$ of \cite{AMM.IM}
is related to the parameter $z$ here as $\xi^2=z$.} $z\to
G^{2/3}(z-2\mathfrak{u})$ (where $G$ and $\mathfrak{u}$ are some
functions of times to be fixed yet) \footnote{As it will be seen
later $u=\mathfrak{u}|_{\tau=0}$ coincides with $u$ appeared earlier
in section \ref{KdV_section}.}
 that generates the change of times to the moment variables,
\be \tilde t_{2m+1}=\frac{1}{(2m+1)G^{\frac{2m+1}{3}}}\oint{
\frac{(v(z)+W(z))dz}{(z-2\mathfrak{u})^{m+\frac{1}{2}}}}
\label{tildetau} \ee Then, the partition functions in old and new
variables are related by the formula (see \cite[sect.3]{AMM.IM} for
the detailed definitions and derivations): \be
Z_K(t)=G^{-\frac{1}{24}}e^{U_{KK}(t)}Z_K(\tilde t) \label{mvars-1}
\ee This formula is correct for any functions $G$ and
$\mathfrak{u}$. We specify them so that \be
\tilde\tau_1=\tilde\tau_3=0,\;\;(\tilde{\tau}_{2k+1}-\frac{1}{3}\delta_{k,1}=\tilde{t}_{2k+1})
\label{mvars-2} \ee Then, \be {\cal
F}^{(0)}=U_{KK}(t)=\frac{1}{2}\oint_\infty\oint_\infty
\rho^{(0|2)}(z_1,z_2|2\mathfrak{u})(\phi(z_1)+\Phi(z_1))(\phi(z_2)+\Phi(z_2))dz_1dz_2\\
{\cal F}^{(1)}=-\frac{1}{24}\log(G) \label{mvars-3} \ee where
$\rho^{(0|2)}(z_1,z_2|2\mathfrak{u})$ is given by formula
(\ref{rho_0_2_g}) and $\phi(z)=\sum\tau_kz^{k+1/2},\,\Phi(z)=\sum
T_kz^{k+1/2},\,2\phi'(z)=v(z),\,2\Phi'(z)=W(z)$

Introduce now, after \cite{AJKM}, the function \be
S(a):=\oint\frac{(v(z)+W(z))dz}{\sqrt{z-2a}}= \sum_{k=0}^\infty \frac{(2k+1)!!}{k!}\,t_{2k+1}\,a^k\ee
Then, from
(\ref{tildetau}) it follows that \be
S(\mathfrak{u})=0\\
S'(\mathfrak{u}):=\frac{\p S(a)}{\p a}\Big|_{a=\mathfrak{u}}=-{G} \label{Gfun} \ee and, for
$k>1$, \be \tilde\tau_{2k+1}=\frac{S^{(k)}}{(2k+1)!!(-S')^{\frac{2k+1}{3}}}, \ee
where $S^{(k)}=S^{(k)}(u)$. Now one can represent the operator $\nabla(z)$ in terms of the
derivative with respect to $\mathfrak{u}$
\be
\nabla(x)=f(x)\Omega+\frac{\p}{\p
\tau}(x)
\ee
\be
f(x)=\frac{1}{(x-2\mathfrak{u})^\frac{3}{2}}\\
\Omega=-\frac{1}{S'}\frac{d}{d \mathfrak{u}}
\ee
where operator $\frac{\p}{\p\tau}(x)$ acts only on explicit dependence of $S^{(k)}$ on $\tau$'s.
Then,
\be \nabla(x){\cal
F}^{(0)}=\oint\rho^{(0|2)}(x,z|2\mathfrak{u})(\phi(z)+\Phi(z))dz-\frac{1}{S'}\frac{1}{(x-2\mathfrak{u})^{\frac{3}{2}}}\frac{\p}{\p \mathfrak{u}}{\cal F}^{(0)}
\ee
The derivative of the two-point function splits
\be
\frac{\p}{\p \mathfrak{u}}\rho^{(0|2)}(z_1,z_2|2\mathfrak{u})=\frac{1}{(z_1-2\mathfrak{u})^{\frac{3}{2}}(z_2-2\mathfrak{u})^{\frac{3}{2}}}
\ee
and
\be \frac{\p}{\p \mathfrak{u}} {\cal
F}^{(0)}\sim\left(\oint\frac{(\phi(x)+\Phi(x))dx}{(x-2\mathfrak{u})^\frac{3}{2}}\right)^2\sim(\tilde\tau_1)^2=0
\ee
In this way one gets the well-known result, which is general for
matrix models: the second derivative of the planar free energy
depends only on ramification points (parameter $\mathfrak{u}$ in our case)
\be
\nabla(z_1)\nabla(z_2){\cal F}^{(0)}=\rho^{(0|2)}(z_1,z_2|2\mathfrak{u})
\ee
Acting again with the operator $\nabla(z_3)$, one gets
\be
\nabla(z_1)\nabla(z_2)\nabla(z_3){\cal F}^{(0)}=-\frac{1}{S'}\frac{1}{((z_1-2\mathfrak{u})(z_2-2\mathfrak{u})(z_3-2\mathfrak{u}))^\frac{3}{2}}
\label{3tochka}
\ee
From the observation that $[\Omega,\nabla(x)]=0$ with help of the
relation
\be
\nabla(z_1)\cdots\nabla(z_k)(S')^{-1}=\Omega^k(S')^{-1}\prod_{i=1}^kf(z_k)
\ee
one derives
\be
\nabla(z_1)\cdots\nabla(z_k)\mathcal{F}^{(0)}=-\left(-\frac{1}{S'}\frac{\p}{\p
\mathfrak{u}}\right)^{k-3}\frac{1}{S'}\prod_{i=1}^kf(z_i)
\ee where, if one
wants to get the expression for zero times $\tau$, one should
substitute
\be S^{(s)}|_{\tau=0}=\sum_{k=s}^N\frac{(2k+1)!!}{(k-s)!}\,T_{2k+1}u^k=M_{s-1}
\label{zerotau}\ee
\begin{equation}
 \mathfrak{u}|_{\tau=0}=u,\;\;\;\sum_{k=0}^N\frac{(2k+1)!!}{k!}\,T_{2k+1}\,u^k=0
\end{equation}
which coincides with the string equation, (\ref{string_eq}). Thus, we obtain the following generalization of (\ref{rho_0_k_g}):
\begin{equation}
 \rho^{(0|k)}(z_1,\cdots,z_k)=-\left(-\frac{1}{M_0}\frac{\p}{\p
u}\right)^{k-3}\frac{1}{M_0}\prod_{i=1}^k\frac{1}{(z_i-2{u})^{3/2}}
\label{rho_0_k_g-gen}
\end{equation}
Let us note that $\frac{\d M_s}{\d u}=(2s+3)M_{s+1}$ and $M_s,\,u$
have a nice description in terms of the curve (see e.g.
(\ref{reduced_curve})). From (\ref{mvars-3}), (\ref{Gfun}) and
(\ref{3tochka}) one gets \be {\cal F}^{(1)}=-\frac{1}{24}\log
S'=-\frac{1}{24}\log\left(\frac{\p^3}{\p\tau_0^3}{\cal
F}^{(0)}\right) \ee For higher genera it is probably too naive to
expect expressions for multi-resolvents to be same simple as for
genera 0 and 1, but one can get (less simple) explicit expressions.
If the variables $u$ and $J_k:=S^{(k)}$ for $k>0$ are considered as
independent, then \be
\frac{\p}{\p \tau} (x)=\sum_{k=1}f^{(k)}(x)\frac{\p}{\p J_k}\\
\frac{d}{d \mathfrak{u}}=\frac{\p }{\p \mathfrak{u}}+\sum_{k=1}J_{k+1}\frac{\p}{\p J_k}
\ee
with $f^{(k)}(x)=\frac{\p^k}{\p \mathfrak{u}^k}f(x)$.
Thus
\be
\nabla(x)=-\frac{f(x)}{J_1}\frac{\p}{\p \mathfrak{u}}+\sum_{k=1}\left(f^{(k)}(x)-\frac{J_{k+1}f(x)}{J_1}\right)\frac{\p}{\p J_k} \label{nabla_u_J}
\ee
This operator is rather easy to apply to low genera, because in terms of the variables $J_k$,
expressions for free energies are simple, for instance,
\be
{\cal F}^{(2)}=\frac{1}{9216}\left(-\frac{J_4}{J_1^3}+\frac{29J_2J_3}{J_1^4}-\frac{7J_2^3}{5J_1^5}\right)
\ee
\be
\nabla(x){\cal F}^{(2)}= \frac{1}{2}\left( f^{(1)}(x)-{\frac {J_{{2}}f(x)}{J_{{1}}}} \right)  \left( {
\frac {1}{192}}\,{\frac {J_{{4}}}{{J_{{1}}}^{4}}}-{\frac {29}{720}}\,{
\frac {J_{{2}}J_{{3}}}{{J_{{1}}}^{5}}}+{\frac {7}{144}}\,{\frac {{J_{{
2}}}^{3}}{{J_{{1}}}^{6}}} \right) +\\
+\frac{1}{2}\left( f^{(2)}(x)-{\frac {J_{{3}}f(x)}{J_{{1}}}} \right)  \left( {\frac {29}{2880}}\,{\frac {J_{{3}}}{{J_
{{1}}}^{4}}}-{\frac {7}{240}}\,{\frac {{J_{{2}}}^{2}}{{J_{{1}}}^{5}}}
 \right)
+{\frac {29J_2}{5760J_1^4}}\, \left( f^{(3)}(x)-{\frac {J_{{4}}f(x)}{
J_{{1}}}} \right)-\\
-{\frac {1}{1152J_1^3}}\, \left( f^{(4)}(x)-{\frac {J_{{5}}f(x)}{J_{{1}}}} \right)
\ee
For the Gaussian branch $M_k=M\delta_{0,k},\;2u=s$, this gives
\be
\rho^{(2|1)}(x)=\nabla(x){\cal F}^{(2)}\Big|_{\tau=0}=\frac{f^{(4)}(x)}{1152\,M^3}=\frac{105}{128\,M^3\,(x-s)^{\frac{11}{2}}}
\ee

One can easily see that, in the generic DV case,
$\rho^{(p|m)}(x_1,\ldots,x_m)$ can be expressed through the coefficients
$M_0,\ldots,M_{N-1}$ and $P(x)$ only. From formula (\ref{nabla_u_J}) it follows that
the singularities in $x_i$ can only be of the form $f^{(k)}(x_i)|_{\tau=0}$, i.e. $1/P^{2k+3}(x_i)$.

Here we list several first densities (for general $N$) (using
variables $u,M_0,\ldots,M_{N-1}$ instead of
$T_1,\ldots,T_{2N+1}$):

\be \rho^{(0|1)}(x)=-W(x)/\sqrt{x}-P(x)\cdot y_r(x) \ee

\be
\rho^{(0|2)}(x_1,x_2)=\frac{1}{(x_1-x_2)^2}\left\{\frac{P^2(x_1)+P^2(x_2)}{P(x_1)P(x_2)}-\frac{x_1+x_2}{\sqrt{x_1x_2}}\right\}
\ee

\be
\rho^{(0|3)}(x_1,x_2,x_3)=\frac{-1}{M_0}\frac{1}{P^3(x_1)P^3(x_2)P^3(x_3)}
\ee

\be
 \rho^{(0|4)}(x_1,x_2,x_3,x_4)=\frac{3}{M_0^2}\,\frac{1}{P^3(x_1)P^3(x_2)P^3(x_3)P^3(x_4)}\cdot \\
\cdot\left(\frac{1}{P^2(x_1)}+\frac{1}{P^2(x_2)}+\frac{1}{P^2(x_3)}+\frac{1}{P^2(x_4)}-\frac{M_1}{M_0}\right)
\ee

\be \rho^{(1|1)}(x)=\frac{-1}{8M_0^2}\frac{M_0-M_1P^2(x)}{P^5(x)}\ee

\ni Note that the dependence on $x$ and $u$ enters only through the
difference $x-2u=P^2(x)$ (analogously to what we had in the $N=1$
case (see formula (\ref{97})), again except for the non-meromorphic
parts in $\rho^{(0|1)}$ and $\rho^{(0|2)}$). This type of solutions
to the Virasoro constraints is sometimes referred to as one-cut
solutions (for the obvious reason).

\subsection{KdV/DV solution and absence of poles}
Taking into account what was said at the end of the previous subsection, one can state that
{\bf if $Z(T)$ satisfies the reduced Virasoro constraints
$\check{L}_{-1},\;\check{L}_{0}$, then
$Z[T]=\tau_{\hbox{\tiny{KP}}}\;\;\Longleftrightarrow
$ the curve $\mathcal{C}_{2,2N-1}$ degenerates to a sphere and
$\rho^{(p|m)}$ for all $p,m$ have no poles at marked points on the sphere which came
from degenerated handles, i.e. at the zeroes of $y_r(x)$.}

Proof of the genus-zero (regarding degeneration of the curve) part of
this conjecture is presented in s.$\ref{sect_eq_to_w}$. It
is actually enough to check the condition for higher $p$ only for
the one-point functions $\rho^{(p|1)}(x)$, because when performing
iterations (\ref{rec_rel}) the singularities just can not arise in
the multi-densities $\rho^{(p|m)}(x_1,..,x_m)$ due to the symmetry
between $x_1,..,x_m$. Then, for each $p$ we have the following
condition: the numerator of $\rho^{(p|1)}(x)$ obtained through
(\ref{rec_rel}), which is a polynomial in $x$, should be
divisible by $y_r(x)$ in the denominator.
In the general case, the remainder is a polynomial of degree $N-2$, and thus
one has $N-1$ equations, which are the first order linear PDEs on $F^{(p)}$.
Hence, the function $\wt$ of $N-1$ variables is completely determined.

The absence of the poles in all $\rho^{(p|1)}$ is equivalent to
vanishing of the integrals

\be \oint_{A_i}\rho^{|1)}(x)dx=0,\hspace{2cm}\forall i=1\ldots N-1
\ee

\ni where the $A$-cycles encircle handles (which are actually pinched):

\begin{center}
\includegraphics[scale=0.7]{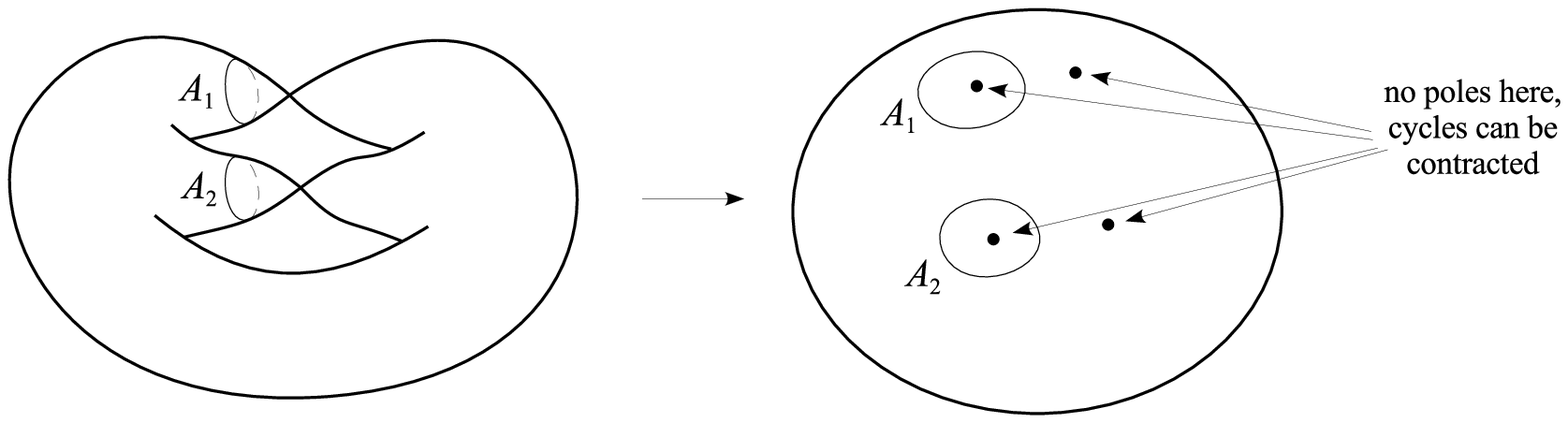}
\end{center}
and $\rho^{|1)}=\sum_{p=0}^{\infty}g^{2p}\rho^{(p|1)}$ is the total
one-point function. Therefore, one can actually forget about this
points and the initial curve, and work in terms of the reduced curve
$P(x)=\sqrt{x-2u}$.

\subsection{A particular case: Chebyshev spectral curves}
There is a special choice of values of $T_k$ which is usually referred to as conformal
background \cite{MSS}. This choice leads to the Chebyshev curves\footnote{In
\cite{Chebyshev_curve} it was shown how the Chebyshev curves emerge in Liouville theory.}.
Indeed, {\bf if values of times $T_k=T_k(\lambda)$ are given by

\be
\sum_{k=0}^NT_{2k+1}\xi^{k+1/2}\frac{(2k+1)!!}{2^{2k-1}}=\frac{2N-1}{\sqrt{2\pi}}(-\lambda)^{N-1/2}\left[K_{N-1/2}(-\lambda/\xi)\xi\right]_+\label{conformal_bgr}\ee

\ni then there exists a solution $u=-{\lambda\over 4}$ of the string equation
(\ref{string_eq}) such that

\be
x=P^2-\frac{1}{2}\lambda=\frac{\lambda}{2}\rT_2(P/\sqrt{\lambda})\ee

\be y=\lambda^{N-1/2}\rT_{2N-1}(P/\sqrt{\lambda})\ee

\ni or, equivalently

\be y=\lambda^{N-1/2}\rT_{N-1/2}\left(\frac{2x}{\lambda}\right) \ee
}
We denote here by $\rT_*$ the Chebyshev polynomials. The choice of times (\ref{conformal_bgr})
defines the conformal background.

In order to prove this claim, one can set $\lambda=1$ without loss of
generality. According to what is said at the end of s.\ref{sect_eq_to_w},
the space of curves of type
(\ref{curve_w_x})-(\ref{curve_whitham}) covers $N$ times the space of
$T_k$'s. Then, it is enough to prove that the times given by the Chebyshev curve
$y=\rT_{N-1/2}(2x)$ are such that (\ref{conformal_bgr}) satisfies. Taking
into account that
$T_{2k+1}=-\frac{1}{2k+1}\res_{x=\infty}\left[x^{-k-1/2}y(x)dx\right]$,
it is correct if
 \be
\sum_{k=-\infty}^N(\rT_{\nu})_k\,\Gamma(k+1/2)\,x^{k-1/2}=(-1)^{\nu}\nu\,K_{\nu}(-1/x)\;\;\;\;\;(\nu\equiv
N-1/2) \ee
is correct. We denote here by $K_*$ the modified Bessel
functions of the second kind and $\rT_{\nu}(x)\equiv\sum_k(\rT_{\nu})_k x^{k-1/2}$. Using
the integral representation for the $\Gamma$-function and deforming the
contour, one can rewrite this equation as
\be \int_{-1}^{+\infty}\rT_{\nu}(\tau)e^{-\tau
l}l\,d\tau=(-1)^{\nu}\nu\,K_{\nu}(-l)\;\;\;\;\;(l\equiv 1/x>0)\ee

\ni Then, integrating by parts in the l.h.s. and using the integral
representation for the Bessel function in the r.h.s., one comes to

\be \int_{-1}^{+\infty}\rT'_{\nu}(\tau)e^{-\tau
l}\,d\tau=\frac{1}{2}\int_Ce^{-\frac{l}{2}(t+\frac{1}{t})}\,\nu
t^{\nu-1}dt \ee

\ni where contour $C$ goes from $0+$ to $+\infty$ encircling $0$ one
time. One can check this equality just by changing the variables
${t+{1}/{t}}=\tau\,,\;t=t_{\pm}=\tau\pm\sqrt{\tau^2-1}$ and using
$\rT_{\nu}(\tau)=(t_+^\nu+t_-^\nu)/2$.

Note that in the general case if one has a curve
\begin{equation}
 \rT_{p}(x)=\rT_q(y)
\end{equation}
\ni then due to the composition property of the Chebyshev
polynomials, $\rT_p\circ \rT_q=\rT_{pq}$ the curve can be parameterized in
the obvious way
\begin{equation}
 x=\rT_q(P)\hspace{2cm}y=\rT_p(P)
\end{equation}

\section{Generalized Kontsevich Model and duality}
\setcounter{equation}{0}

In this section we review some basic aspects of GKM.
The next section contains some more detailed analysis of the $p=3$ case and
includes finding the curve, the first densities and action of the $p-q$ duality on them.

\subsection{Generalized Kontsevich Model}
Partition function of GKM is given by the following matrix integral (\cite{GKM}):
\begin{equation}
 Z(t)=\frac{1}{\mathcal{N}(A)}\int dXe^{\frac{1}{g^2}\mathrm{Tr}\left\{-\frac{X^{p+1}}{p+1}+A^pX\right\}}
\end{equation}
It is a function of the Miwa variables
\begin{equation}
 T_k+\tau_k\equiv t_{k}=\frac{p}{p+1}\delta_{p+1,k}+\frac{1}{k}\mathrm{Tr}A^{-k}
\end{equation}
Actually it depends only on times $t_k,\,k\neq 0\,(\mathrm{mod}\,p)$.
The Ward identities in this case are the $\mathcal{W}$-constraints (\cite{FKN1,GKM})
\begin{equation}
 \mathcal{W}^{(k)}_nZ=0,\;k=2..p,\;n\geqslant -k+1 \label{W-constraints}
\end{equation}
For $k=2$ these are the Virasoro constraints
$\mathcal{W}^{(2)}_n=\mathcal{L}_n$. One can define the densities
analogously to the $p=2$ case
\begin{equation}
 \rho^{(p|m)}(z_1,\ldots,z_m)=\D(z_1)\cdots\D(z_2)\F^{(p)}|_{\tau=0},\;\;\;\;\D(z)=\sum_{k=0}^{\infty}\frac{1}{z^{k/p+1}}\frac{\d}{\d \tau_{k}}
\end{equation}
In particular,
\begin{equation}
 \rho^{(0|1)}(x)=-W(x)+y(x),\;\;W(x)=\sum_{k=1}^{p+q}kx^{\frac{k}{p}-1}T_k
\end{equation}
and one expects $y(x)$ to satisfy an algebraic equation
\begin{equation}
 P_{p,q}(y,x)=0
\end{equation}
defining the spectral curve. The polynomial $P_{p,q}(y,x)$ is of degree $p$ in $y$ and
of degree $q$ in $x$. The GKM is known to be relevant for describing the $(p,q)$-model of $2d$ gravity
when the first $p+q$ times are turned on ($T_n=0,\;n>p+q$).
The well known solution to (\ref{W-constraints}) is given by the $\tau$-function of the
$p$-reduced KP-hierarchy (\cite{Douglas,FKN1}). The spectral curve in this case degenerates to
the rational one. When one further specifies to the conformal background the curve has
the form of \cite{Chebyshev_curve}
\begin{equation}
 T_p(y)=T_q(x).
\end{equation}

\subsection{p-q duality}
The $p-q$ duality is obvious when one formulates theory in terms of the Douglas equation
\cite{Douglas} generalizing (\ref{Douglas})
\begin{equation}
 [{\bf L},{\bf M}]=2g
\end{equation}
where ${\bf L}$ and ${\bf M}$ are differential operators of orders $p$ and $q$ respectively.
However, when one formulates the theory in terms of $p$-reduced KP-hierarchy (where $L$ plays
role of the Lax operator) or in terms of the matrix model, the duality becomes implicit.
In \cite{FKN} an explicit change of time variables $T\leftrightarrow \bar{T}$ of the hierarchy
which connects $(p,q)$ and $(q,p)$ models was found. In \cite{KM} the $p-q$ duality was considered
from the point of view of the GKM.

Within our approach, one first of all expects that the duality relates the functions
$\tilde{\tilde{F}}^{\pq{p}{q}}$ and $\tilde{\tilde{F}}^{\pq{q}{p}}$
(generalizing those considered in s.\ref{sect_RVC}) which describe
ambiguities in the solutions of the $\mathcal{W}$-constrains (so that fixing them is equivalent to
fixing the solution\footnote{Still, fixing the solution is equivalent to fixing
periods of all the multi-densities.}). They both depend on a certain number (equal to the genus
of the spectral curve) of combinations of times $T$, consequently the change of times should
map the set of these combinations for the $(p,q)$ model to that of the $(q,p)$ model.
Then (after identification of $\tilde{\tilde{F}}^{\pq{p}{q}}$ and
$\tilde{\tilde{F}}^{\pq{q}{p}}$) one expects that the curves for the two models should coincide,
that is, there is an isomorphism $\mathcal{C}_{\pq{p}{q}}\stackrel{\phi}\longrightarrow
\mathcal{C}_{\pq{q}{p}}$:
\begin{equation}
 P_{p,q}(y^{\pq{p}{q}},x^{\pq{p}{q}})\sim P_{{q},{p}}(\phi^*y^{\pq{q}{p}},\phi^*x^{\pq{q}{p}})
\end{equation}
In simple cases, this is just the interchange
$x\leftrightarrow y$: $\phi^*y^{\pq{q}{p}}=x^{\pq{p}{q}},\,\phi^*x^{\pq{q}{p}}=y^{\pq{p}{q}}$.
Note that, as soon as we talk about solutions to the $\mathcal{W}$-constrains only, without
further specifying them as KP solutions,
in variance with \cite{FKN,quas,KM} we deal with a more general, ``off-KP" duality.

\section{Generalized Kontsevich matrix model $\mathbf{(p=3)}$}
\setcounter{equation}{0}

\subsection{Some general formulae and definitions}

Most of what comes up below is quite analogous to the $p=2$ case,
hence, we present it very briefly (one can find some parallel
consideration in \cite{Kr}).

\be\mathcal{W}^{(2)}_n\equiv\mathcal{L}_n=\frac{1}{3}\left(\frac{1}{2g^2}\sum_{i+j=-3n}ijt_it_j+\sum_{i-j=-3n}it_i\frac{\d}{\d
t_j}+\frac{g^2}{2}\sum_{i+j=3n}\frac{\d^2}{\d t_i\d
t_j}+\frac{1}{3}\delta_{n,0}\right)\ee

\ba
\mathcal{W}^{(3)}_n=\frac{1}{9g}\left(\frac{1}{3g^2}\sum_{p+q+r=-3n}pqrt_pt_qt_r+
\sum_{p+q-r=-3n}pqt_pt_q\frac{\d}{\d t_r}\right.\nn\\
\left.+g^2\sum_{p-q-r=-3n}pt_p\frac{\d^2}{\d t_q\d t_r}
+\frac{g^4}{3}\sum_{p+q+r=3n}\frac{\d^3}{\d t_p \d t_q\d t_r}
\right) \ea

\be \mathcal{L}_nZ=0,\;\;\;n\from -1 \ee

\be \mathcal{W}^{(3)}_nZ=0,\;\;\;n\from -2 \ee

\be t_{k}=0,\;\;\; k=0\;(\mathrm{mod}\;3)\ee

\noindent
The shift of times: $t_k=\tau_k+T_k,$ $T_k\neq 0$ for $k=1\ldots (q+3)$. Thus,
it can be referred to as (3,q) model.

\be \D_i(x):=\sum_{k=0}^{\infty}\frac{1}{x^{k+1}}\frac{\d}{\d
\tau_{3k+i}}\hspace{15mm}\;\;i=1,2 \ee

\be
v_i(x)=\sum_{k=0}^{\infty}(3k+i)x^{k}\tau_{3k+i}\hspace{15mm}W_i(x)=\sum_{k=0}^{1+\left[\frac{q-i}{3}\right]}(3k+i)x^{k}T_{3k+i}\;\;\;\;i=1,2\ee

\be \rho_i(x):=\D_i(x)\F \;\;\;i=1,2 \ee

\noindent
The loop equations are
\ba
2g^1\sum_{n=-1}^\infty\frac{1}{x^{n+1}}\mathcal{W}^{(2)}_n\,Z=2(\tau_1+T_1)(\tau_2+T_2)+\frac{g^2}{3x}+\rho_1(x)\rho_2(x)+\frac{g^2}{2}(\D_1(x)\rho_2(x)+\D_2(x)\rho_1(x))+\nn\\
+P^-_x\left\{(u_1(x)+W_1(x))\rho_1(x)+(u_2(x)+W_2(x))\rho_2(x)\right\}=0
\ea

\ba
9g^3\sum_{n=-2}^\infty\frac{1}{x^{n+2}}\mathcal{W}^{(3)}_n=\frac{g^6}{3}\D_1^3+\frac{g^6}{3x}\D^3_2+g^4P^-_x
\left\{\frac{(u_1+W_1)}{x}\D_2^2+(u_2+W_2)\D_1^2\right\}+\nn\\
+g^2P^-_x
\left\{\frac{(u_1+W_1)^2}{x}\D_2+(u_2+W_2)^2\D_1\right\}+\frac{8}{3}(\tau_2+T_2)^3+4(\tau_1+T_1)^2(\tau_4+T_4)+\frac{(\tau_1+T_1)^3}{3x}\\
=\frac{1}{3}P_{\left\{x^{-n}|n\in\mathbb{Z}_+\right\}}\left\{x\,:\left(g^2\D-(u+W)\right)^3:\right\}\hspace{2cm}\nn
\ea

\be \check{\D}_i(x)=\sum_{k=0}^N\frac{1}{x^{k+1}}\frac{\d}{\d
T_{3k+i}}\ee

\be
\D(x)=\D_1(x)x^{-1/3}+\D_2(x)x^{-2/3}\;\;\;\rho(x)=\rho_1(x)x^{-1/3}+\rho_2(x)x^{-2/3}=\D(x)\F\ee

\be W(x)=W_1(x)x^{-2/3}+W_2(x)x^{-1/3}\ee

\be F[T_1,\ldots,T_{3+q}]=\F|_{\tau=0}\ee

\be
\rho^{(p|m)}(z_1,\ldots,z_m)=\D(z_1)\cdots\D(z_2)\F^{(p)}|_{\tau=0}
\ee

\subsection{Some first densities}
Let
\be
\rho^{(0|1)}_1={-W_2+y_2x^{1/3}}\;\;\;\;\;\rho^{(0|1)}_2={-W_1+y_1x^{2/3}}\label{rho_0_1_p_3}
\ee

\be \rho^{(0|1)}(x)={-W(x)+y(x)}\ee

\ni where
\be y_1+y_2=y\ee
\ni From the definition of $\rho^{(0|1)}$ one has

\be
T_{3k+i}=-\frac{1}{3k+i}\res_{x=\infty}\left\{x^{-k-i/3}y(x)dx\right\}\;\;,\;\;\left.\frac{\d
\F^{(0)}}{\d
\tau_{3k+i}}\right|_{\tau=0}=-\res_{x=\infty}\left\{x^{k+i/3}y(x)dx\right\}\ee

Substituting\footnote{What follows is actually true only for $q<7$.} (\ref{rho_0_1_p_3}) into the loop equations for genus zero
one obtain the following equation for $y$ defining the
$\mathcal{C}_{\pq{p=3}{q}}$ curve:

\be y^3+3A(x)y+B(x)=0\ee

\ni where $A$ and $B$ are the following \textit{polynomials}:

\be
A(x)=-P^+_x\left\{W_1(x)W_2(x)+\left[W_1(x)(\check{\D}_1(x)F^{(0)})+W_2(x)(\check{\D}_2(x)F^{(0)})\right]\right\}/x
\ee

\be
B(x)=-P^+_x\left\{W_1^3/x+W_2^3+3\left[W_1(\check{\D}_2F^{(0)})^2/x+W_2(\check{\D}_1F^{(0)})^2\right]+\right. \\
\left.+3\left[W_1^2(\check{\D}_2F^{(0)})/x+W_2^2(\check{\D}_1F^{(0)})\right]\right\}/x
\ee

There is also the following identity:
 \be \hspace{2cm}y_1y_2=-A\ee
which, together with $y_1+y_2=y$, allows one to express
$y_i$ through $y$ and $A$.

Some formulae which are useful for computing the higher densities are

\be
\D_i(x)P_z^-\{u_i(z)h(z)\}=-3x^{\frac{i}{3}}\d_xx^{-\frac{i}{3}}\left\{\frac{zP_x^-h(x)-xP_z^-h(z)}{z-x}\right\}+P_z^-\{u_i(z)\D_i(x)h(z)\}\ee

\be y'=-\frac{1}{3}\frac{B'+3A'y}{y^2+A}\ee

An explicit expression for $\rho^{(0|2)}$ in a certain special case can
be found in s.\ref{sect_dual_rho_0_2}.

\subsection{Solving reduced constraints and (3,2)$\leftrightarrow$(2,3)
duality}\label{dual_red}

Now consider reduced Virasoro constraints for $q=2$ coming from
$\mathcal{W}^{(3)}_{-2}$, $\mathcal{L}_{-1}$, $\mathcal{L}_{0}$-constraints
accordingly:

\be \frac{4}{3}T_2^3+2{T_1^2T_4}+10T_2T_5\frac{\d F}{\d
T_1}+8T_4^2\frac{\d F}{\d T_2}+\frac{5^2}{2}T_5^2\frac{\d F}{\d
T_4}=0\ee

\be 2{T_1T_2}+4T_4\frac{\d F}{\d T_1}+5T_5\frac{\d F}{\d T_2}=0\ee

\be \frac{g^2}{3}+T_1\frac{\d F}{\d T_1}+2T_2\frac{\d F}{\d
T_2}+4T_4\frac{\d F}{\d T_4}+5T_5\frac{\d F}{\d T_5}=0\ee

\noindent
The general solution is as follows

\ba F[T_1,T_2,T_4,T_5]=-\frac{g^2}{15}\log T_5- {\displaystyle \frac
{1}{5}} \, {\displaystyle \frac {{T_{1}}\,{T_{2}}^{2}}{{T_{5}}}} +
{\displaystyle \frac {4}{75}} \,{\displaystyle \frac {{T_{2}}^{3}
\,{T_{4}}}{{T_{5}}^{2}}}  - {\displaystyle \frac {2}{25}} \,
{\displaystyle \frac {{T_{4}}^{2}\,{T_{1}}^{2}}{{T_{5}}^{2}}}  +
{\displaystyle \frac {16}{125}} \,{\displaystyle \frac {{T_{4}}^{3
}\,{T_{1}}\,{T_{2}}}{{T_{5}}^{3}}}  - {\displaystyle \frac {32}{
625}} \,{\displaystyle \frac {{T_{4}}^{4}\,{T_{2}}^{2}}{{T_{5}}^{
4}}} \nn \\
\mbox{} - {\displaystyle \frac {128}{9375}} \,{\displaystyle \frac
{{T_{4}}^{6}\,{T_{1}}}{{T_{5}}^{5}}}  + {\displaystyle \frac
{512}{46875}} \,{\displaystyle \frac {{T_{4}}^{7}\,{T_{2}}
}{{T_{5}}^{6}}}  - {\displaystyle \frac {4096}{5859375}} \,
{\displaystyle \frac {{T_{4}}^{10}}{{T_{5}}^{8}}} + \wt
\left({\displaystyle \frac {125\,{T_{1}}\,{T_{5}}^{3} - 100\,{T_{2}}
\,{T_{4}}\,{T_{5}}^{2} +
16\,{T_{4}}^{4}}{5^4\cdot3^{4/5}\cdot2^{-5/5}\cdot{(-T_{5})}^{(16/5)}}}
\right) \ea

\ni where $\wt$ is an arbitrary function. This is the (3,2) model.
The (2,3) model corresponds to the $N^{\hbox{\tiny{$(p=2)$}}}=2$
case considered in subsections \ref{sect_RVC} and \ref{ell_rat}.
Let us denote
$\bar{T}_1=T_1^{\hbox{\tiny{(N=2)}}},\bar{T}_3=T_3^{\hbox{\tiny{(N=2)}}},\bar{T}_5=T_5^{\hbox{\tiny{(N=2)}}}$.

Using the transformations\footnote{The variable $T_3$ can be considered as auxiliary. All (3,2)
quantities are actually independent on it.
For instance, one can choose $T_3=\frac{4}{15}\frac{T_4^2}{T_5}$ so that the last two terms
in the r.h.s. of (\ref{u32-u23}) cancel each other.} (the same as in \cite{FKN}):

\be {\bar{T}_{1}}={T_{1}} - {\displaystyle \frac {4}{5}} \,
{\displaystyle \frac {{T_{2}}\,{T_{4}}}{{T_{5}}}}  - {\displaystyle
\frac {9}{20}} \,{\displaystyle \frac {{T_{3}}^{2} }{{T_{5}}}}  +
{\displaystyle \frac {18}{25}} \,{\displaystyle \frac
{{T_{3}}\,{T_{4}}^{2}}{{T_{5}}^{2}}}  - {\displaystyle \frac
{4}{25}} \,{\displaystyle \frac {{T_{4}}^{4}}{{T_{5}}^{3}} } \ee \be
\,{\bar{T}_{3}}={T_{3}} - {\displaystyle \frac {4}{5}} \,
{\displaystyle \frac {{T_{4}}^{2}}{{T_{5}}}} \ee \be
\,{\bar{T}_{5}}= - {\displaystyle \frac {2}{3}} \,{T_{5}} \! \ee

\ni one can see that

\be \eta_0 \sim
\frac{\bar{T}_1\bar{T}_5-\frac{3}{10}\bar{T}_3^2}{\bar{T}_5^{6/5}}\sim\frac
{125\,{T_{1}}\,{T_{5}}^{3} - 100\,{T_{2}} \,{T_{4}}\,{T_{5}}^{2} +
16\,{T_{4}}^{4}}{{T_{5}}^{(16/5)}}\ee

\ni and, thus, the arbitrary functions in the two models can be identified:
$\wt_{\hbox{\tiny{(2,3)}}}=\wt_{\hbox{\tiny{(3,2)}}}$. Moreover, one
gets the same relation as in \cite{FKN}:

\be u_{\hbox{\tiny{(3,2)}}}[T_1,T_2,T_4,T_5]=
u_{\hbox{\tiny{(2,3)}}}[\bar{T}_1,\bar{T}_3,\bar{T}_5]+{\displaystyle
\frac {2}{25}} \,{\displaystyle \frac {{T_{4}}^{2} }{{T_{5}}^{2}}} -
{\displaystyle \frac {3}{10}} \, {\displaystyle \frac
{{T_{3}}}{{T_{5}}}}\label{u32-u23}\ee

\be u_{\hbox{\tiny{(3,2)}}}=\frac{\d^2 F_{\hbox{\tiny{(3,2)}}}}{\d
T_1^2}\hspace{2cm} u_{\hbox{\tiny{(2,3)}}}=\frac{\d^2
F_{\hbox{\tiny{(2,3)}}}}{\d \bar{T}_1^2}\ee

\ni Note that we used so far the $W$ constraints only and did not assume that $Z$ is a
$\tau$-function of the KP-hierarchy. Therefore, this is an off-KP duality.

\subsection{Duality between (2,3) and (3,2) curves}

For $N^{\hbox{\tiny{(p=3)}}}=1$ (i.e. the $(3,2)$-model) we have the
curve

\be y^3+3A(x)y+B(x)=0\ee

\be A(x)=-20T_4T_5\cdot x-(5T_1T_5+8T_2T_4)\ee
\be B(x)=-125T_5^3\cdot x^2 -(150T_2T_5^2+64T_4^3)\cdot
x-(60T_2^2T_5+48T_1T_4^2+75T_5^2\frac{\d
F_{\hbox{\tiny{(3,2)}}}^{(0)}}{\d T_1})\ee
This is an elliptic curve. For $N^{\hbox{\tiny{(p=2)}}}=2$
(i.e. the $(2,3)$-model) we had the following elliptic curve:
\be y^2-25\bar{T}_5^2\cdot x^3-30\bar{T}_3\bar{T}_5\cdot
x^2-(10\bar{T}_1\bar{T}_5+9\bar{T}_3^2)\cdot
x-(6\bar{T}_1\bar{T}_3+10\bar{T}_5\frac{\d
F_{\hbox{\tiny{(2,3)}}}^{(0)}}{\d \bar{T}_1})=0\ee

After the change of variables $\bar{T}\rightarrow T$ and the identification
$\wt_{\hbox{\tiny{(2,3)}}}=\wt_{\hbox{\tiny{(3,2)}}}$ described in
s.\ref{dual_red}, one can check that the $j$-invariants of the curves
coincide with each other:

\be j^{\hbox{\tiny{(2,3)}}}=j^{\hbox{\tiny{(3,2)}}} \ee

\ni i.e. the curves are isomorphic. At the critical points, the
isomorphism is given just by

\be  x^{\hbox{\tiny{(2,3)}}}\sim y^{\hbox{\tiny{(3,2)}}}
\;\;\;\;\;y^{\pq{2}{3}}\sim x^{\hbox{\tiny{(3,2)}}}\ee

Further details on equivalence of these Riemann surfaces can be found in Appendix B.

\ni Note that this is again the off-KP duality.

\subsection{Duality between $\rho^{(0|1)}$'s}

The isomorphism between the elliptic curves
$\mathcal{C}_{\pq{3}{2}}\stackrel{\phi}\longrightarrow
\mathcal{C}_{\pq{2}{3}}$ is given at the level of coordinate functions by the
following linear (!) explicit expressions

\be \phi^*y^{\pq{2}{3}}=\frac{10}{3}T_5\cdot
x^{\pq{3}{2}}+\frac{4}{5}\frac{T_4}{T_5}\cdot
y^{\pq{3}{2}}+2\left(T_2+\frac{32}{75}\frac{T_4^3}{T_5^2}\right) \ee

\be \phi^*x^{\pq{2}{3}}=+\frac{1}{5}\frac{1}{T_5}\cdot
y^{\pq{3}{2}}+\frac{3}{5}\frac{T_3}{T_5} \ee

\ni Using this, one can see that

\be
\phi^*\left[\rho^{(0|1)}_{\pq{2}{3}}\right]_{\tt}=\left[\rho^{(0|1)}_{\pq{3}{2}}\right]_{\tt}\;\;(\hspace{-3mm}\mod
exact\;form) \ee

\ni where we regard $\rho^{(0|1)}=\rho^{(0|1)}(x)dx$ as a 1-form,
and $\left[\rho^{(0|1)}\right]_{\tt}$ is its meromorphic part
(without the $W(x)$-term, which simply cancels the singular part of
the expansion of $y$ w.r.t. $x$ nearby $x=\infty$ and thus, given
the local coordinate $x$, can be recovered easily). Therefore, the
meromorphic differentials $y^{\pq{3}{2}}dx^{\pq{3}{2}}$ and
$y^{\pq{2}{3}}dx^{\pq{2}{3}}$ differ by a meromorphic differential
with zero periods. This can be also expressed as

\be \int\limits_{A_i}\rho^{(0|1)}_{\pq{3}{2}} =
\int\limits_{\phi(A_i)}\rho^{(0|1)}_{\pq{2}{3}}
\hspace{1cm}\int\limits_{B_i}\rho^{(0|1)}_{\pq{3}{2}} =
\int\limits_{\phi(B_i)}\rho^{(0|1)}_{\pq{2}{3}} \ee

\ni where $A_i,B_i$ are the cycles on the $\mathcal{C}_{\pq{3}{2}}$
(we assume that the contours do not encircle the ramification points
$x=0$ and $x=\infty$ produced by $W(x)$-terms in $\rho^{(0|1)}$ in
both cases).

\subsection{On $\rho^{(0|2)}$ in (2,3) and (3,2) models}\label{sect_dual_rho_0_2}
The isomorphism $\phi$, of course, connects the holomorphic
differentials on the curves:

\be
\phi^*\,\frac{dx^{\pq{2}{3}}}{y^{\pq{2}{3}}}\sim\frac{dx^{\pq{3}{2}}}{\left(y^{\pq{3}{2}}\right)^2+A}
\ee
For the sake of simplicity, in what follows we work at the critical point (only $T_1$ and $T_5$ are
switched on). Then one has
\be A=-5T_1T_5\;\;\;\;\;\;B=125T_5^3x^2+75T_5^2\frac{\d F}{\d
T_1}\ee

\be \phi^*\,\frac{dx^{\pq{2}{3}}}{y^{\pq{2}{3}}}=-2\cdot 5 \cdot
T_5\,\frac{dx^{\pq{3}{2}}}{\left(y^{\pq{3}{2}}\right)^2+A} \ee
In the $p=2$ case, there is a decomposition of
$\rho^{(0|2)}$
\begin{equation}
\rho^{(0|2)}=\rho_{hol}^{(0|2)}+\rho_{glob}^{(0|2)}-\rho_{loc}^{(0|2)}
\end{equation}

\be \rho^{(0|2)}_{\pq{3}{2},hol}=5^4\,T_5^4\,\frac{\d^2F}{\d
T_1^2}\frac{dxdz}{(y^2(x)+A)(y^2(z)+A)} \ee

\[
\rho_{\pq{3}{2},glob}^{(0|2)}=\frac{dxdz}{4(x-z)^2(y^2(x)+A)(y^2(z)+A)}\left[2\cdot
3\cdot 5^2 \cdot T_1^2T_5^2+2\cdot 3\cdot 5\cdot T_1T_5\cdot
y(x)y(z)+\right.
\]
\be\left.-y(x)\left\{2^23^25^2T_5^2\frac{\d F}{\d
T_1}+5^3T_5^3(z^2+2xz)\right\}-y(z)\left\{2^23^25^2T_5^2\frac{\d
F}{\d T_1}+5^3T_5^3(x^2+2xz)\right\}  \right] \ee

\be
\rho_{\pq{3}{2},loc}^{(0|2)}=\frac{\left[x^{1/3}(2z+x)+z^{1/3}(2x+z)\right]dx\,dz}{4\,x^{2/3}\,z^{2/3}\,(x-z)^2}
\ee
Note that $\rho^{(0|2)}_{\pq{3}{2},glob},\;\rho^{(0|2)}_{\pq{3}{2},hol},\;\rho^{(0|2)}_{\pq{3}{2},loc}$
have the same defining properties as $\rho^{(0|2)}_{\pq{2}{3}}$ discussed in s.\ref{Resolvents_section}.
One can also try to formulate it in
terms of the correlators (see s.\ref{CFT_section}), however, here this is more difficult to find an
appropriate local CFT. The scheme of cuts of
the covering
$\mathcal{C}_{\pq{3}{2}}\stackrel{\pi_{\pq{3}{2}}}{\longrightarrow}\mathbb{CP}^1$
is

\begin{center}
\includegraphics[scale=0.6]{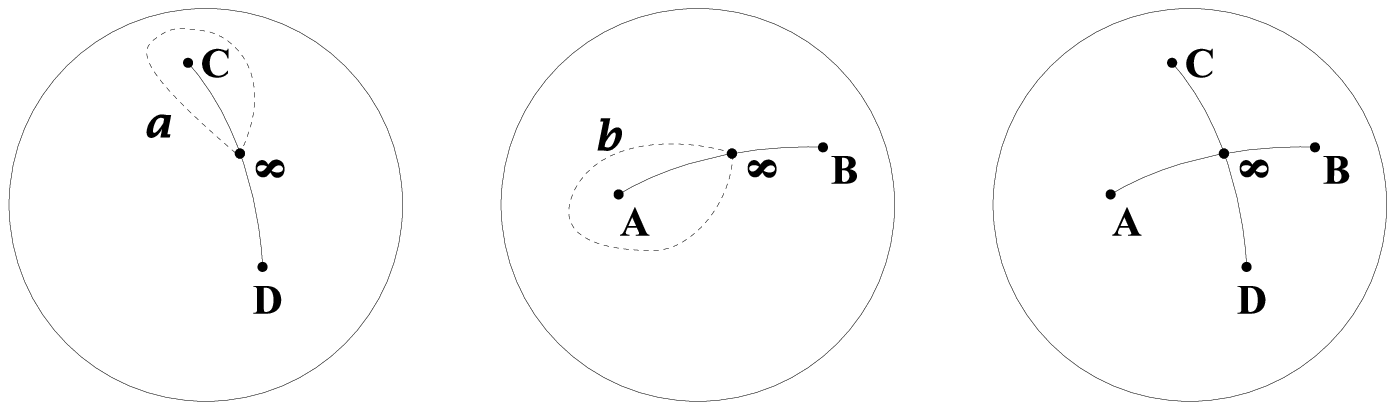}
\end{center}

\ni Here $[a]$ and $[b]$ are the standard homology basis on the
torus. If one considers, say, the collection of fields
$(X_1,X_2,X_3)$ on the sheets, there is no linear combination
of them which diagonalizes the monodromies around all points
simultaneously.

For the (2,3) model one has

\be
\rho_{\pq{2}{3},hol}^{(0|2)}=\frac{5^2}{2^2}\,\bar{T}_5^2\,\frac{\d^2F}{\d
\bar{T}_1^2}\frac{dxdz}{y(x)y(z)} \ee

\be
\rho_{\pq{2}{3},glob}^{(0|2)}=\frac{dxdz}{(x-z)^2y(x)y(z)}\left[\frac{1}{16}\left\{5^2\bar{T}_5^2xz+10\bar{T}_1\bar{T}_5\right\}(x+z)+5\bar{T}_5\frac{\d
F}{\d \bar{T}_1}\right] \ee

\be
\rho_{\pq{2}{3},loc}^{(0|2)}=\frac{\left[x+z\right]dx\,dz}{4\,x^{1/2}\,z^{1/2}\,(x-z)^2}
\ee

\ni The singularity at $x=z$:

\be
\rho_{\pq{3}{2},glob}^{(0|2)}\sim\rho_{\pq{3}{2},loc}^{(0|2)}\sim\frac{3}{2}\frac{dxdz}{(x-z)^2}
\ee

\be
\rho_{\pq{2}{3},glob}^{(0|2)}\sim\rho_{\pq{2}{3},loc}^{(0|2)}\sim\frac{1}{2}\frac{dxdz}{(x-z)^2}
\ee

Both $\rho^{(0|2)}_{\pq{2}{3}}$ and $\rho^{(0|2)}_{\pq{3}{2}}$
can be constructed in terms of the coverings
$\mathcal{C}_{\pq{2}{3}}\stackrel{\pi_{\pq{2}{3}}}{\longrightarrow}\mathbb{CP}^1$
and
$\mathcal{C}_{\pq{3}{2}}\stackrel{\pi_{\pq{3}{2}}}{\longrightarrow}\mathbb{CP}^1$.
The curves are isomorphic
$\mathcal{C}_{\pq{3}{2}}\stackrel{\phi}\rightarrow
\mathcal{C}_{\pq{2}{3}}$ but, of course,
$\pi_{\pq{3}{2}}\neq\pi_{\pq{2}{3}}\phi$.

\section*{Acknowledgements}

Our work is partly supported by Russian Federal Nuclear Energy
Agency, by the joint grant 06-01-92059-CE, by NWO project
047.011.2004.026, by INTAS grant 05-1000008-7865, by l'Agence
Nationale de la Recherche under the grants ANR-05-BLAN-0029-01 and
ANR-06-BLAN-3 137168 (A.A.), by the Russian President's Grants of
Support for the Scientific Schools NSh-3035.2008.2
(A.A.,A.Mir.,A.Mor.) and NSh-3036.2008.2 (P.P.), by RFBR grants
06-02-17383 (A.A.), 07-02-00878 (A.Mir. and P.P.) and 07-02-00645
(A.Mor.). In addition, the research of A.A. was supported by
European RTN under the contract 005104 "ForcesUniverse" and by
Dynasty foundation.

\appendix

\section{Derivation of (\ref{w_res})} \label{app_disp_KdV}

\begin{lemma}
 There exist the following formulae for $T_k$ and $\frac{\d F^{(0)}}{\d T_k}:$
\begin{equation}
 T_{k}=-\frac{1}{k}\res_{P=\infty}\left\{P\,M L^{-k/2}d P\right\}
\end{equation}
\begin{equation}
 \frac{\d F^{(0)}}{\d T_{k}}=-\res_{P=\infty}\left\{P\,ML^{k/2}d P\right\}
\end{equation}
\label{lemma_w_res}
\end{lemma}
\proof Let us denote

\be H:=\sum_{k=0}^N\frac{(2k+1)!!}{k!}\,T_{2k+1}\,u^k \ee

\ni Then, one can rewrite (\ref{str}) as $\frac{\d H}{\d T_1}=0$.
From (\ref{dc}) and the dispersionless counterpart of (\ref{dc2}) it
follows that

\be \frac{\d^2 F^{(0)}}{\d T_1\d T_{2k+1}}=-\res
L^{k+1/2}=\frac{(2k+1)!!}{(k+1)!}u^{k+1}\label{res_Ftt}\ee

\be \Rightarrow\;\;\frac{\d u}{\d
T_{2k+1}}=\frac{(2k+1)!!}{k!}u^{k}\frac{\d u}{\d T_1}\ee

\ni Then

\be \frac{\d H}{\d T_1}=0 \Rightarrow  \frac{\d H}{\d
T_{2k+1}}=0 \ee

\ni thus $H=0$ (note that we have also checked in the way that this
equation respects all the KdV-flows). One can see explicitly that
the equation $H=0$ can be represented as follows

\be T_1=-\res\left\{P\,ML^{-1/2}d P\right\} \ee

\ni For $k>0$ the equations \be
T_{2k+1}=-\frac{1}{2k+1}\res\left\{P\,ML^{-(2k+1)/2}d P\right\} \ee

\ni follow straightforwardly from formula (\ref{Q_oper}) for $M$.
Thus, it remains to verify that

\be \frac{\d F^{(0)}}{\d
T_{2k+1}}=-\res\left\{P\,ML^{k+1/2}d P\right\}=\sum_{s=1}^N(-1)^s M_{s-1}u^{s+k-1}\frac{(2k+1)!!\cdot(2s-1)!!}{(s+k+1)!}\ee

\ni Differentiating it w.r.t. $T_1$ (note that, doing this, we lose
nothing, because in the KdV hierarchy only the quantities
containing, at least, first derivatives of $F$ in $T_1$ make any
sense) and taking into account that, from the string equation
(\ref{str}), one obtains \be \frac{\d M_{s-1}}{\d
T_1}=(2s+1)M_s\frac{\d u}{\d
T_1}\hspace{1cm},\hspace{1cm}M_0\frac{\d u}{\d T_1}=-1\ee

\ni one finally arrives at \be \frac{\d F^{(0)}}{\d T_{2k+1}\d
T_1}=\frac{(2k+1)!!}{(k+1)!}u^{k+1}\ee

\ni which is the same as (\ref{res_Ftt}). $\square$

For reference, we also write down explicit formulae for $T_k$, $M_s$
and $\frac{\d F^{(0)}}{\d T_{k}}$:

\be
T_{2k+1}=\frac{1}{2k+1}\sum_{s=k}^N(-1)^{s-k}M_{s-1}u^{s-k}\frac{(2s-1)!!}{(2k-1)!!\cdot(s-k)!}
\label{change_T_Y} \ee

\begin{equation}
 M_s=\sum_{k=1}^NT_{2k+1}\frac{(2k+1)!!}{(2s+1)!!(k-s-1)!}u^{k-s-1}
\end{equation}

\be \frac{\d F^{(0)}}{\d
T_{2k+1}}=\sum_{s=1}^N(-1)^sM_{s-1}u^{s+k-1}\frac{(2k+1)!!\cdot(2s-1)!!}{(s+k+1)!}
\ee

\section{On equivalence of Riemann surfaces}

\subsection{$y^2=x^2-1$ and other hyperelliptic curves}

The complex curve $y^2=x^2-1$ is an ordinary Riemann sphere with
coordinate $z$, where $x = \frac{1}{2}(z+1/z)$ and $y =
\frac{1}{2}(z-1/z)$. There are no holomorphic differentials on it,
and the kernel, the bi-differential with a double pole on the
diagonal can be represented in a variety of ways: \be B(z_1,z_2) =
\frac{dz_1dz_2}{(z_1-z_2)^2} = \frac{1}{2}
\frac{dx_1dx_2}{(x_1-x_2)^2} \left(1 +
\frac{x_1x_2-1}{y_1y_2}\right) = \frac{1}{2}
\frac{dy_1dy_2}{(y_1-y_2)^2}\left(1+\frac{y_1y_2 +1}{x_1x_2}\right)
\ee

\noindent
For the generic hyperelliptic curve, $y^2 = P_n(x) = a_n x^n + \ldots
+ a_0$ \be B = \frac{1}{2}\frac{dx_1dx_2}{(x_1-x_2)^2} \left(1 +
\frac{\tilde P_n(x_1,x_2)}{y_1y_2}\right) \ee where the new
polynomial $\tilde P_n$ is defined by the three conditions: $\tilde
P_n(x,x) = P_n(x)$, symmetricity $\tilde P_n(x_1,x_2) = \tilde
P_n(x_2,x_1)$ and restricted growth at infinity, $\tilde
P_n(x_1,x_2) \sim x_1^{[(n+1)/2]}$. This means that for even $n=2m$
\be \tilde P_{2m}(x_1,x_2) = a_{2m}x_1^mx_2^m +
a_{2m-1}\frac{x_1^mx_2^{m-1} + x_1^{m-1}x_2^m}{2}
+ \\
+a_{2m-2}
\Big(\alpha_{2m-2}\frac{x_1^mx_2^{m-2} + x_1^{m-2}x_2^m}{2} +
(1-\alpha_{2m-2})x_1^{m-1}x_2^{m-1}\Big) + \ldots
\ee
while for odd $n=2m-1$
$$
\tilde P_{2m-1}(x_1,x_2) =
a_{2m-1}\frac{x_1^{m}x_2^{m-1}+x_1^{m-1}x_2^{m}}{2} + \ldots
$$
Extra $\frac{m(m-1)}{2}$ $\alpha$-parameters describe possible
bilinear combinations of holomorphic differentials which can be
added to $B$, for example, $\alpha_{2m-2}$ is actually a coefficient
in front of $x_1^{m-2}x_2^{m-2}(x_1-x_2)^2$. This ambiguity can be
fixed by specifying $A$-periods of the kernel $B$. In particular, the
Bergmann kernel corresponds to the trivial A-periods
\be
\oint_{A_i} B=0
\ee

\subsection{$y^3 = x^2-1$ }

This curve is an ordinary torus with extended discrete $Z_2\times
Z_3$ symmetry. In the dual coordinates $X= y$, $Y=x$ it acquires the
usual hyperelliptic form $Y^2 = X^3+1$, and the (un-normalized)
holomorphic differential and the Bergmann kernel are \be v =
\frac{dX}{Y},\ \ \ B =
\frac{1}{2}\frac{dX_1dX_2}{(X_1-X_2)^2}\left(1 +
\frac{X_1^2X_2+X_1X_2^2+2}{2Y_1Y_2}\right) \ee In coordinates $x,y$
one has instead a representation as a triple covering with the three
ramification points at $x=\pm 1$ and $x = \infty$. The local coordinates
in the vicinities of these points are \be
\begin{array}{cc|c}
x = \pm 1 + \xi^3, & y \sim \xi; dx \sim \xi^2d\xi\\
x = \frac{1}{\xi^3}, & y \sim \frac{1}{\xi^2}; dx \sim
\frac{d\xi}{\xi^4}
\end{array}
\ee Accordingly, the holomorphic differential and the Bergmann kernel
acquire the form \be v = \frac{2}{3} \frac{dx}{y^2}, \ \ \ B =
\frac{dx_1dx_2}{(x_1-x_2)^2}
\left(\frac{y_1^2+y_1y_2+y_2^2}{3y_1y_2}\right)^2 -
\frac{1}{9}\frac{(y_1+y_2)dx_1dx_2}{y_1^2y_2^2} \ee In the first
term in $B$ the numerator vanishes when $y_2 = \pm e^{2\pi i/3}y_1$
and cancels the unwanted poles at the points $x_1=x_2$ with $y_1\neq y_2$. The
second term serves to cancel poles at infinity.

The modified Begrmann kernel, with the pole at $x_2=-x_1$ is \be B^* =
\frac{1}{2}\frac{dX_1dX_2}{(X_1-X_2)^2}\left(1 -
\frac{X_1^2X_2+X_1X_2^2+2}{2Y_1Y_2}\right) = \\
=
-\frac{dx_1dx_2}{(x_1+x_2)^2}
\left(\frac{y_1^2+y_1y_2+y_2^2}{3y_1y_2}\right)^2 +
\frac{1}{9}\frac{(y_1+y_2)dx_1dx_2}{y_1^2y_2^2} \ee

\subsection{$y^3 = A_1(x)y + C_2(x)$ from $Y^2 = P_3(X)$}

Here $A_1(x) = a_{11}x + a_{10}$, \ $C_2(x) = c_{22}x^2 + c_{21}x +
c_{20}$ and $P_3(X) = p_{33}X^3 + p_{32}X^2 + p_{31}X + p_{30}$.

The equation $y^3 = A_1(x)y + C_2(x)$ is quadratic in $x$, and the function
$x(y)$ has the four order-two ramification points at $y=\infty$ and at the
three roots of discriminant in the expression
\be x = \frac{-(c_{21}+a_{11}y) \pm
\sqrt{(c_{21}+a_{11}y)^2 - 4c_{22}(c_{20} + a_{10}y -
y^3)}}{2c_{22}} \ee At infinity \be X \sim \frac{1}{\xi^2}, \ \ \ Y
\sim \frac{1}{\xi^3} \ee and \be x \sim \frac{1}{\xi^3}, \ \ \ y
\sim \frac{1}{\xi^2} \ee Therefore, \be Y = ux+ vy + w, \ \ \ X = py
+q \ee The holomorphic differential $\frac{dX}{Y} \sim
\frac{dy}{ux+vy+w}$ does not have poles at zeroes of the
denominator, provided these zeroes are located exactly at the ramification
points, where $y$ has double zeroes. This means that actually \be Y
= ux+vy+w \sim 2c_{22}x+a_{11}y+c_{21} \ee Substituting this
expression into $Y^2 = P_3(X)$ one obtains
$$
\left(p_{33}p^3\right)y^3 + \left(3p_{33}p^2q + p_{32}p^2 -
a_{11}^2\right)y^2 + \left(3p_{33}pq^2 + 2p_{32}pq + p_{31}p -
2a_{11}c_{21}\right)y - 4a_{11}c_{22}xy =
$$ $$ = 4c_{22}^2x^2 + 4c_{22}c_{21}x + c_{21}^2 - P_3(q)
$$
which coincides with $y^3 = A_1(x)y + C_2(x)$ provided
$$
\left\{
\begin{array}{c}
2a_{11}^2 = p^2P_3''(q) \\
p_{33} p^3 = 4c_{22} \\
P_3(q) = c_{21}^2-4c_{20}c_{22} \\
2a_{11}c_{21} - 4a_{10}c_{22} = p P_3'(q)
\end{array}
\right.
$$
These equations can be used to define $p,q$ and $A_1(x)$ for given
$P_3(x)$. The remaining freedom is $x \rightarrow \alpha x + \beta$ and
it can be used to fix two coefficients in $C_2(x)$.

\end{document}